\documentclass[12 pt,twoside,aps,prd,amsmath,amssymb,
tightenlines,showpacs,showkeys,eqsecnum]{revtex4}

\usepackage{graphicx}
\usepackage{mathptmx}
\usepackage{bm}
\newcommand {\ve}{\varepsilon}
\newcommand {\pr}{\partial}

\newcommand {\bg}{\bar \gamma}
\newcommand {\G}{\Gamma}
\newcommand {\bp}{\bar \psi}
\newcommand {\p}{\psi}

\def\myfigure #1#2#3#4
{\begin{figure}[ht]\begin{center}
\includegraphics[width=#2 \textwidth]{#1.eps}
\parbox[t]{#4\textwidth}{\caption{#3}\label{#1}}
\end{center}\end{figure}}

\def \myfigures #1#2#3#4#5#6#7#8
{\begin{figure}[ht]
    \begin{center}
        \includegraphics[width=#2 \textwidth]{#1.eps}
        \hfill
        \includegraphics[width=#6 \textwidth]{#5.eps}
        \parbox[t]{#4\textwidth}{\caption {#3}\label{#1}}
        \hfill
        \parbox[t]{#8\textwidth}{\caption {#7}\label{#5}}
    \end{center}
\end{figure} }

\begin{document}
\baselineskip -24pt
\title{Nonlinear Spinor Fields and its role in Cosmology}
\author{Bijan Saha}
\affiliation{Laboratory of Information Technologies\\
Joint Institute for Nuclear Research\\
141980 Dubna, Moscow region, Russia} \email{bijan@jinr.ru}
\homepage{http://bijansaha.narod.ru}

\begin{abstract}

Different characteristic of matter influencing the evolution of the
Universe has been simulated by means of a nonlinear spinor field.
Exploiting the spinor description of perfect fluid and dark energy
evolution of the Universe given by an anisotropic Bianchi type-VI,
VI$_0$, V, III, I or isotropic Friedmann-Robertson-Walker (FRW) one
has been studied. It is shown that due to some restrictions on
metric functions, initial anisotropy in the models Bianchi type-VI,
VI$_0$, V and III does not die away, while the anisotropic Bianchi
type-I models evolves into the isotropic one.

\end{abstract}



\maketitle

\bigskip

\section{Introduction}

Being related to almost all stable elementary particles such as
proton, electron and neutrino, spinor field, especially Dirac
spin-$1/2$ play a principal role at the microlevel. However, in
cosmology, the role of spinor field was generally considered to be
restricted. Only recently, after some remarkable works  by different
authors (e.g., Henneaux, 1980; Saha, 2001a, 2004a, 2004b, 2006c,
2006d; Armend$\acute a$riz-Pic$\acute o$n and Greene, 2003; Ribas
{\it et al}, 2005; Souza and Kremer, 2008), showing the important
role that spinor fields play on the evolution of the Universe, the
situation began to change. This change of attitude is directly
related to some fundamental questions of modern cosmology: (i)
problem of initial singularity; (ii) problem of isotropization and
(iii) late time acceleration of the Universe.

{\bf (i) Problem of initial singularity:} One of the problems of
modern cosmology is the presence of initial singularity, which means
the finiteness of time. The main purpose of introducing a nonlinear
term in the spinor field Lagrangian is to study the possibility of
the elimination of initial singularity. In a number of papers it was
shown that the introduction of nonlinear spinor field into the
system indeed gives rise to singularity-free models of the Universe
(Saha, 2001a, 2001b, 2004a), (Saha and Shikin, 1997a, 1997b).

{\bf  (ii) problem of isotropization:} Although the Universe seems
homogenous and isotropic at present, it does not necessarily mean
that it is also suitable for a description of the early stages of
the development  of the Universe and there are no observational data
guaranteeing the isotropy in the era prior to the recombination. In
fact, there are theoretical arguments that support the existence of
an anisotropic phase that approaches an isotropic one (Misner 1968).
The observations from Cosmic Background Explorer's differential
radiometer have detected and measured cosmic microwave background
anisotropies in different angular scales. These anisotropies are
supposed to hide in their fold the entire history of cosmic
evolution dating back to the recombination era and are being
considered as indicative of the geometry and the content of the
universe. More about cosmic microwave background anisotropy is
expected to be uncovered by the investigations of microwave
anisotropy probe. There is widespread consensus among the
cosmologists that cosmic microwave background anisotropies in small
angular scales have the key to the formation of discrete structure.
It was found that the introduction of nonlinear spinor field
accelerates the isotropization process of the initially anisotropic
Universe (Saha, 2001a, 2004a, 2006c).

{\bf  (iii) late time acceleration of the Universe:} Detection and
further experimental reconfirmation of current cosmic acceleration
pose to cosmology a fundamental task of identifying and revealing
the cause of such phenomenon. This fact can be reconciled with the
theory if one assumes that the Universe id mostly filled with
so-called dark energy. This form of matter (energy) is not
observable in laboratory and it does not interact with
electromagnetic radiation. These facts played decisive role in
naming this object. In contrast to dark matter, dark energy is
uniformly distributed over the space, does not intertwine under the
influence of gravity in all scales and it has a strong negative
pressure of the order of energy density. Based on these properties,
cosmologists have suggested a number of dark energy models, those
are able to explain the current accelerated phase of expansion of
the Universe. In this connection a series of papers appeared
recently in the literature, where a spinor field was considered as
an alternative model for dark energy (Ribas {\it et al}, 2005; Saha,
2006d, 2006e, 2007).

It should be noted that most of the works mentioned above were
carried out within the scope of Bianchi type-I cosmological model.
Results obtained using a spinor field as a source of Bianchi type-I
cosmological field can be summed up as follows: A suitable choice of
spinor field nonlinearity\\

(i) {\it accelerates the isotropization process} (Saha, 2001a,
2004a,
2006c);\\

(ii) {\it gives rise to a singularity-free Universe} (Saha, 2001a,
2004a,
2004b, 2006c);\\

(iii) {\it generates late time acceleration} (Ribas {\it et al},
2005; Saha, 2006d;  Souza and Kremer, 2008).

Given the role that spinor field can play in the evolution of the
Universe, question that naturally pops up is, if the spinor field
can redraw the picture of evolution caused by perfect fluid and dark
energy, is it possible to simulate perfect fluid and dark energy by
means of a spinor field? Affirmative answer to this question was
given in the a number of papers (Krechet {\it et al}, 2008; Saha,
2010a, 2010b). In those papers the authors have shown that different
types of perfect fluid and dark energy can be described by nonlinear
spinor field. In (Saha, 2010a) we used two types of nonlinearity,
one occurs as a result of self-action and the other resulted from
the interaction between the spinor and scalar field. It was shown
that the case with induced nonlinearity is the partial one and can
be derived from the case with self-action. In (Saha, 2010b, 2011) we
give the description of generalized Chaplygin gas and modified
quintessence in terms of spinor field and study the evolution of the
Universe filled with nonlinear spinor field within the scope of a
Bianchi type-I and FRW cosmological model. The purpose of this paper
is to extend that study within the framework of other Bianchi
models.

\section{Simulation of perfect fluid with nonlinear spinor field}

Nonlinear quantum Dirac fields were used by Heisenberg (1953, 1957)
in his ambitious unified theory of elementary particles. They are
presently the object of renewed interest since the widely known
paper by Gross and Neveu (1974). A nonlinear spinor field, suggested
by the symmetric coupling between nucleons, muons, and leptons, has
been investigated by Finkelstein {\it et al.} (1951) in the
classical approximation.

In this section we simulate different types of perfect fluid and
dark energy by means of a nonlinear spinor field.

\subsection{Spinor field Lagrangian}

For a spinor field $\p$, the symmetry between $\p$ and $\bp$ appears
to demand that one should choose the symmetrized Lagrangian (Kibble,
1961). Keeping this in mind we choose the spinor field Lagrangian as
(Saha, 2001a):
\begin{equation}
L_{\rm sp} = \frac{i}{2} \biggl[\bp \gamma^{\mu} \nabla_{\mu} \psi-
\nabla_{\mu} \bar \psi \gamma^{\mu} \psi \biggr] - m_{\rm sp}\bp
\psi + F, \label{lspin}
\end{equation}
where the nonlinear term $F$ describes the self-action of a spinor
field and can be presented as some arbitrary functions of invariant
generated from the real bilinear forms of a spinor field. For
simplicity we consider the case when $F = F(S)$ with $S = \bp \psi$.
Here $\nabla_\mu$ is the covariant derivative of spinor field:
\begin{equation}
\nabla_\mu \psi = \frac{\partial \psi}{\partial x^\mu} -\G_\mu \psi,
\quad \nabla_\mu \bp = \frac{\partial \bp}{\partial x^\mu} + \bp
\G_\mu, \label{covder}
\end{equation}
with $\G_\mu$ being the spinor affine connection. In \eqref{lspin}
$\gamma$'s are the Dirac matrices in curve space-time and obey the
following algebra
\begin{equation}
\gamma^\mu \gamma^\nu + \gamma^\nu \gamma^\mu = 2 g^{\mu\nu}
\label{al}
\end{equation}
and are connected with the flat space-time Dirac matrices $\bg$ in
the following way
\begin{equation}
 g_{\mu \nu} (x)= e_{\mu}^{a}(x) e_{\nu}^{b}(x) \eta_{ab},
\quad \gamma_\mu(x)= e_{\mu}^{a}(x) \bg_a, \label{dg}
\end{equation}
where $\eta_{ab}= {\rm diag}(1,-1,-1,-1)$ and $e_{\mu}^{a}$ is a set
of tetrad 4-vectors. The spinor affine connection matrices $\G_\mu
(x)$ are uniquely determined up to an additive multiple of the unit
matrix by the equation
\begin{equation}
\nabla_\mu \gamma_\nu = \frac{\pr \gamma_\nu}{\pr x^\mu} -
\G_{\nu\mu}^{\rho}\gamma_\rho - \G_\mu \gamma_\nu + \gamma_\nu
\G_\mu = 0, \label{afsp}
\end{equation}
with the solution
\begin{equation}
\Gamma_\mu = \frac{1}{4} \bg_{a} \gamma^\nu \partial_\mu e^{(a)}_\nu
- \frac{1}{4} \gamma_\rho \gamma^\nu \Gamma^{\rho}_{\mu\nu},
\label{sfc}
\end{equation}

Varying \eqref{lspin} with respect to $\bp (\psi)$ one finds the
spinor field equations:
\begin{subequations}
\label{speq}
\begin{eqnarray}
i\gamma^\mu \nabla_\mu \psi - m_{\rm sp} \psi + \frac{dF}{dS} \psi &=&0, \label{speq1} \\
i \nabla_\mu \bp \gamma^\mu +  m_{\rm sp} \bp - \frac{dF}{dS} \bp
&=& 0, \label{speq2}
\end{eqnarray}
\end{subequations}
The energy-momentum tensor of the spinor field is given by
\begin{equation}
T_{\mu}^{\rho}=\frac{i}{4} g^{\rho\nu} \biggl(\bp \gamma_\mu
\nabla_\nu \psi + \bp \gamma_\nu \nabla_\mu \psi - \nabla_\mu \bar
\psi \gamma_\nu \psi - \nabla_\nu \bp \gamma_\mu \psi \biggr) \,-
\delta_{\mu}^{\rho} L_{\rm sp} \label{temsp}
\end{equation}
where $L_{\rm sp}$ in account of spinor field equations
\eqref{speq1} and \eqref{speq2} takes the form
\begin{equation}
L_{\rm sp} = - S \frac{dF}{dS} + F(S). \label{lsp}
\end{equation}

We consider the case when the spinor field depends on $t$ only. In
this case for the components of energy-momentum tensor we find
\begin{subequations}
\begin{eqnarray}
T_0^0 &=& m_{\rm sp}S - F, \label{t00s}\\
T_1^1 = T_2^2 = T_3^3 &=& S \frac{dF}{dS} - F. \label{t11s}
\end{eqnarray}
\end{subequations}
A detailed study of nonlinear spinor field was carried out in Saha
(2001a, 2004a, 2006c). In what follows, exploiting the equation of
states we find the concrete form of $F$ which describes various
types of perfect fluid and dark energy.

\subsection{perfect fluid with a barotropic equation of state}

First of all let us note that one of the simplest and popular model
of the Universe is a homogeneous and isotropic one filled with a
perfect fluid with the energy density $\ve = T_0^0$ and pressure $p
= - T_1^1 = -T_2^2 = -T_3^3$ obeying the barotropic equation of
state
\begin{equation}
p = W \ve, \label{beos}
\end{equation}
where $W$ is a constant. Depending on the value of $W$ \eqref{beos}
describes perfect fluid from phantom to ekpyrotic matter, namely
\begin{subequations}
\label{zeta}
\begin{eqnarray}
W &=& 0, \qquad \qquad \qquad {\rm (dust)},\\
W &=& 1/3, \quad \qquad \qquad{\rm (radiation)},\\
W &\in& (1/3,\,1), \quad \qquad\,\,{\rm (hard\,\,Universe)},\\
W &=& 1, \quad \qquad \quad \qquad {\rm (stiff \,\,matter)},\\
W &\in& (-1/3,\,-1), \quad \,\,\,\,{\rm (quintessence)},\\
W &=& -1, \quad \qquad \quad \quad{\rm (cosmological\,\, constant)},\\
W &<& -1, \quad \qquad \quad \quad{\rm (phantom\,\, matter)},\\
W &>& 1, \quad \qquad \quad \qquad{\rm (ekpyrotic\,\, matter)}.
\end{eqnarray}
\end{subequations}
The barotropic model of perfect fluid is used to study the evolution
of the Universe. Most recently the relation \eqref{beos} is
exploited to generate a quintessence in order to explain the
accelerated expansion of the Universe (Saha, 2005; Zlatev 1999).

In order to describe the matter given by \eqref{zeta} with a spinor
field let us now substitute $\ve$ and $p$ with $T_0^0$ and $-T_1^1$,
respectively. Thus, inserting $\ve = T_0^0$ and $p = - T_1^1$ from
\eqref{t00s} and \eqref{t11s} into \eqref{beos} we find
\begin{equation}
S \frac{dF}{dS} - (1+W)F + m_{\rm sp} W S= 0, \label{eos1s}
\end{equation}
with the solution (Saha, 2010a, 2010b, 2011)
\begin{equation}
F = \lambda S^{1+W} + m_{\rm sp}S. \label{sol1}
\end{equation}
Here $\lambda$ is an integration constant. Taking into account that
the energy density should be non-negative, we conclude that
$\lambda$ is a negative constant, we write the components of the
energy momentum tensor
\begin{subequations}
\begin{eqnarray}
T_0^0 &=& \nu S^{1+W}, \label{t00sf}\\
T_1^1 = T_2^2 = T_3^3 &=& - \nu W S^{1+W}, \label{t11sf}
\end{eqnarray}
\end{subequations}
where $\lambda = - \nu$, with $\nu$ being a positive constant. As
one sees, the energy density $\ve = T_0^0$ is always positive, while
the pressure $p = - T_1^1 = \nu W S^{1+W}$ is positive for $W
> 0$, i.e., for usual fluid and negative for $W < 0$, i.e. for dark
energy.

In account of it the spinor field Lagrangian now reads
\begin{equation}
L_{\rm sp} = \frac{i}{2} \biggl[\bp \gamma^{\mu} \nabla_{\mu} \psi-
\nabla_{\mu} \bar \psi \gamma^{\mu} \psi \biggr] - \nu S^{1+W},
\label{lspin1}
\end{equation}
Thus a massless spinor field with the Lagrangian \eqref{lspin1}
describes perfect fluid from phantom to ekpyrotic matter. Here the
constant of integration $\nu$ can be viewed as constant of
self-coupling. A detailed analysis of this study was given in
Krechet (2008).

\subsection{Chaplygin gas}

An alternative model for the dark energy density was used by
Kamenshchik {\it et al.} (2001), where the authors suggested the use
of some perfect fluid but obeying "exotic" equation of state. This
type of matter is known as {\it Chaplygin gas}. The fate of density
perturbations in a Universe dominated by the Chaplygin gas, which
exhibit negative pressure was studied by Fabris {\it et al.} (2002).
Model with Chaplygin gas was also studied in the Refs. (Dev {\it et
al}, 2003; Bento {\it et al}, 2002).

Let us now generate a Chaplygin gas by means of a spinor field. A
Chaplygin gas is usually described by a equation of state
\begin{equation}
p = -A/\ve^\gamma. \label{chap}
\end{equation}
Then in case of a massless spinor field for $F$ one finds
\begin{equation}
\frac{(-F)^\gamma d(-F)}{(-F)^{1+\gamma} - A} = \frac{dS}{S},
\label{eqq}
\end{equation}
with the solution (Saha, 2010a, 2010b, 2011)
\begin{equation}
-F = \bigl(A + \lambda S^{1+\gamma}\bigr)^{1/(1+\gamma)}.
\label{chapsp}
\end{equation}
On account of this for the components of energy momentum tensor we
find
\begin{subequations}
\begin{eqnarray}
T_0^0 &=& \bigl(A + \lambda S^{1+\gamma}\bigr)^{1/(1+\gamma)}, \label{edchapsp}\\
T_1^1 = T_2^2 = T_3^3 &=& A/\bigl(A + \lambda
S^{1+\gamma}\bigr)^{\gamma/(1+\gamma)}. \label{prchapsp}
\end{eqnarray}
\end{subequations}
As was expected, we again get positive energy density and negative
pressure. Analogical results were obtained in (Cai {\it et al}, 2008
).

Thus the spinor field Lagrangian corresponding to a Chaplygin gas
reads
\begin{equation}
L_{\rm sp} = \frac{i}{2} \biggl[\bp \gamma^{\mu} \nabla_{\mu} \psi-
\nabla_{\mu} \bar \psi \gamma^{\mu} \psi \biggr] - \bigl(A + \lambda
S^{1+\gamma}\bigr)^{1/(1+\gamma)}. \label{lspin2}
\end{equation}
Setting $\gamma = 1$ we find the result obtained in Saha (2010a).

\subsection{Modified quintessence}

Finally, we simulate modified quintessence with a nonlinear spinor
field. It should be noted that one of the problems that face models
with dark energy is that of eternal acceleration. One of the
possible way to avoid this problem is to introduce a negative
$\Lambda$-term together with a quintessence, which gives rise to a
oscillatory mode of expansion (Cardenas {\it et. al.}, 2003, Saha,
2006a) In order to get rid of that problem quintessence with a
modified equation of state was proposed which is given by (Saha,
2006b)
\begin{equation}
p =  W (\ve - \ve_{\rm cr}), \quad W \in (-1,\,0), \label{mq}
\end{equation}
Here $\ve_{\rm cr}$ some critical energy density.  Setting $\ve_{\rm
cr} = 0$ one obtains ordinary quintessence. It is well known that as
the Universe expands the (dark) energy density decreases. As a
result, being a linear negative function of energy density, the
corresponding pressure begins to increase. In case of an ordinary
quintessence the pressure is always negative, but for a modified
quintessence as soon as $\ve$ becomes less than the critical one,
the pressure becomes positive.

Inserting $\ve = T_0^0$ and $p = - T_1^1$ into \eqref{mq} we find
\begin{equation}
F = - \nu S^{1+W} + m_{\rm sp}S - \frac{W}{1+W}\ve_{\rm cr},
\label{Fmq}
\end{equation}
with $\eta$ being a positive constant. On account of this for the
components of energy momentum tensor we find
\begin{subequations}
\begin{eqnarray}
T_0^0 &=& \nu S^{1+W} + \frac{W}{1+W}\ve_{\rm cr}, \label{edmq}\\
T_1^1 = T_2^2 = T_3^3 &=& - \nu  W S^{1+W} + \frac{W}{1+W}\ve_{\rm
cr}. \label{prmq}
\end{eqnarray}
\end{subequations}

Lagrangian for spinor field describing perfect fluid and modified
quintessence can be written in the following way (Saha, 2011)
\begin{equation}
L_{\rm sp} = \frac{i}{2} \biggl[\bp \gamma^{\mu} \nabla_{\mu} \psi-
\nabla_{\mu} \bar \psi \gamma^{\mu} \psi \biggr] - \nu S^{1+W}
-\frac{W}{1+W}\ve_{\rm cr}. \label{lspin2n}
\end{equation}
One can easily verify, in case of $\ve_{\rm cr} = 0$ \eqref{lspin2n}
corresponds to a perfect fluid, while nontrivial $\ve_{\rm cr}$ with
$W \in (-1,\,0)$ generates modified quintessence. It should be noted
that the restriction $W > -1$ is very important, as it does not
allow the system to cross over phantom divide barrier.

We see that a nonlinear spinor field with specific type of
nonlinearity can substitute perfect fluid and dark energy, thus give
rise to a variety of evolution scenario of the Universe.

\section{Cosmological models with a spinor field}

In the previous section we showed that the perfect fluid and the
dark energy can be simulated by a nonlinear spinor field. In the
section II the nonlinearity was the subject to self-action. In
(Saha, 2010a) we have also considered the case when the nonlinearity
was induced by a scalar field. It was also shown the in our context
the results for induced nonlinearity is some special cases those of
self-interaction. Taking it into mind we study the evolution an
Universe filled with a nonlinear spinor field given by the
Lagrangian \eqref{lspin}, with the nonlinear term $F$ is given by
\eqref{sol1}, \eqref{chapsp} and \eqref{Fmq}.

\subsection{Bianchi type anisotropic cosmological model}

Let us study the evolution of an anisotropic Bianchi type
cosmological model filled with spinor field. In this report we
consider Bianchi type-VI, VI$_0$, V, III, I and FRW models.

We choose the Bianchi type-VI cosmological model in the form (Saha,
2004b)
\begin{equation}
ds^2 = dt^2 - a_1^2 e^{-2mz} dx^2 - a_2^2 e^{2nz} dy^2 - a_3^2 dz^2,
\label{bvi}
\end{equation}
with $a_1,\,a_2,\,a_3$ being the function of $t$ only.  A suitable
choice of $m,\,n$ as well as the metric functions $a_1,\,a_2,\,a_3$
in the BVI given by \eqref{bvi} evokes the following Bianchi-type
universes:
\begin{itemize}
\item
for $m = n$ the BVI metric transforms to a Bianchi-type VI$_0$
(BVI$_0$) one, i.e., $m = n$, BVI $\Longrightarrow$ BVI$_0$ $\in$
open FRW with the line elements
\begin{equation}
ds^2 = dt^2 - a_1^{2} e^{-2mz}\,dx^{2} - a_2^{2} e^{2mz}\,dy^{2} -
a_3^{2}\,dz^2; \label{bvi0}
\end{equation}
\item
for $m = - n$ the BVI metric transforms to a Bianchi-type V (BV)
one, i.e., $m = n$, BVI $\Longrightarrow$ BV $\in$ open FRW with the
line elements
\begin{equation}
ds^2 = dt^2 - a_1^{2} e^{2mz}\,dx^{2} - a_2^{2} e^{2mz}\,dy^{2} -
a_3^{2}\,dz^2; \label{bv}
\end{equation}

\item
for $n = 0$ the BVI metric transforms to a Bianchi-type III (BIII)
one, i.e., $n = 0$, BVI $\Longrightarrow$ BIII with the line
elements
\begin{equation}
ds^2 = dt^2 - a_1^{2} e^{-2mz}\,dx^{2} - a_2^{2} \,dy^{2} -
a_3^{2}\,dz^2; \label{biii}
\end{equation}
\item
for $m = n = 0$ the BVI metric transforms to a Bianchi-type I (BI)
one, i.e., $m = n = 0$, BVI $\Longrightarrow$ BI with the line
elements
\begin{equation}
ds^2 = dt^2 - a_1^{2} \,dx^{2} - a_2^{2} \,dy^{2} - a_3^{2}\,dz^2;
\label{bi}
\end{equation}
\item
for $m=n=0$ and equal scale factor in all three directions the BVI
metric transforms to a Friedmann-Robertson-Walker (FRW) universe,
i.e., $m = n = 0$ and $a=b=c$, BVI $\Longrightarrow$ FRW with the
line elements.
\begin{equation}
ds^2 = dt^2 - a^{2} \bigl(dx^{2} + \,dy^{2} + \,dz^2\bigr).
\label{frw}
\end{equation}
\end{itemize}

Let us go back to the metric Bianchi-type VI. The metric \eqref{bvi}
possesses following nontrivial covariant and contravariant
components:
\begin{eqnarray}
g_{00} = 1, \quad g_{11} = a_1^2 e^{-2mz}, \quad g_{22} = a_2^2
e^{2nz}, \quad g_{33} = a_3^2, \nonumber\\
\nonumber \\
g^{00} = 1, \quad g^{11} =\frac{1}{a_1^2 e^{-2mz}}, \quad g^{22} =
\frac{1}{a_2^2 e^{2nz}}, \quad g^{33} = \frac{1}{a_3^2}. \nonumber
\end{eqnarray}
The nontrivial Christoffel symbols for \eqref{bvi} are
\begin{eqnarray}
\G_{01}^{1} &=& \frac{\dot{a_1}}{a_1},\quad \G_{02}^{2} =
\frac{\dot{a_2}}{a_2},\quad
\G_{03}^{3} = \frac{\dot{a_3}}{a_3}, \nonumber\\
\G_{11}^{0} &=& a_1 \dot{a_1} e^{-2mz},\quad \G_{22}^{0} = a_2
\dot{a_2} e^{2nz},\quad
\G_{33}^{0} = a_3 \dot{a_3},\label{Chrysvi}\\
\G_{31}^{1} &=& -m,\quad \G_{32}^{2} = n,\quad \G_{11}^{3} = \frac{m
a_1^2}{a_3^2} e^{-2mz},\quad \G_{22}^{3} = -\frac{n a_2^2}{a_3^2}
e^{2nz}. \nonumber
\end{eqnarray}

In view of \eqref{dg} we choose the tetrad as follows:
\begin{equation}
e_0^{(0)} = 1, \quad e_1^{(1)} = a_1 e^{-mz}, \quad e_2^{(2)} = a_2
e^{nz}, \quad e_3^{(3)} = a_3. \label{tetradvi}
\end{equation}
From
\begin{equation} \gamma_\mu = e^{(a)}_\mu \bg_a.
\label{gammabg}
\end{equation}
one now finds
\begin{equation}
\gamma_0 = \bg_0, \quad  \gamma_1 = a_1 e^{-mz} \bg_1, \quad
\gamma_2 = a_2 e^{nz} \bg_2, \quad \gamma_3 = a_3 \bg_3.
\label{gbgvi}
\end{equation}
Taking into account that in our case
$$\bg^0 = \bg_0, \quad \bg^1 = -\bg_1, \quad \bg^2 = -\bg_2, \quad
\bg^3 = -\bg_3,$$ one also finds
\begin{equation}
\gamma^0 = \bg^0, \quad  \gamma^1 = \frac{e^{mz}}{a_1} \bg^1, \quad
\gamma^2 = \frac{e^{-nz}}{a_2} \bg^2, \quad \gamma^3 = \frac{1}{a_3}
\bg^3. \label{gbgviup}
\end{equation}

Now we are ready to compute spinor affine connections using
\eqref{sfc} which in our particular case gives
\begin{subequations}
\label{sacvi}
\begin{eqnarray}
\Gamma_0 &= & \frac{1}{4} \bg_{a} \gamma^\nu \partial_t e^{(a)}_\nu
- \frac{1}{4} \gamma_\rho \gamma^\nu \Gamma^{\rho}_{0\nu} = 0,\label{G0vi}\\
\Gamma_1 &=&  - \frac{1}{4} \gamma_\rho \gamma^\nu
\Gamma^{\rho}_{1\nu} =  \frac{1}{2}\Bigl(\dot a_1 \bg^1\bg^0 -
m\frac{a_1}{a_3} \bg^1\bg^3\Bigr) e^{-mz}, \label{G1vi}\\
\Gamma_2 &=&  - \frac{1}{4} \gamma_\rho \gamma^\nu
\Gamma^{\rho}_{2\nu} =  \frac{1}{2}\Bigl(\dot a_2 \bg^2\bg^0 +
n\frac{a_2}{a_3} \bg^2\bg^3\Bigr) e^{nz}, \label{G2vi}\\
\Gamma_3 &= & \frac{1}{4} \bg_{a} \gamma^\nu \partial_z e^{(a)}_\nu
- \frac{1}{4} \gamma_\rho \gamma^\nu \Gamma^{\rho}_{3\nu} =
\frac{1}{2} \dot a_3 \bg^3\bg^0. \label{G3vi}
\end{eqnarray}
\end{subequations}

From \eqref{sacvi} one finds
\begin{equation}
\gamma^\mu \Gamma_\mu = - \frac{1}{2} \Bigl(\frac{\dot a_1}{a_1} +
\frac{\dot a_2}{a_2} + \frac{\dot a_3}{a_3}\Bigr) \bg^0 +
\frac{m-n}{2a_3} \bg^3. \label{gGvi}
\end{equation}

Choosing $m,\,n$ we can thus write the spinor affine connections for
other Bianchi type metrics.

\subsubsection{Bianchi type-VI anisotropic cosmological model}

Let us now study the evolution of the Universe given by a Bianchi
type-VI cosmological model. The Einstein equations corresponding to
the metric \eqref{bvi} take the form:

\begin{subequations}
\label{einbvi}
\begin{eqnarray}
\frac{\ddot a_2}{a_2} +\frac{\ddot a_3}{a_3} +\frac{\dot
a_2}{a_2}\frac{\dot
a_3}{a_3} - \frac{n^2}{a_3^2} &=& \kappa T_{1}^{1}, \label{11bvi}\\
\frac{\ddot a_3}{a_3} +\frac{\ddot a_1}{a_1} +\frac{\dot
a_3}{a_3}\frac{\dot
a_1}{a_1} - \frac{m^2}{a_3^2} &=& \kappa T_{2}^{2}, \label{22bvi} \\
\frac{\ddot a_1}{a_1} +\frac{\ddot a_2}{a_2} +\frac{\dot
a_1}{a_1}\frac{\dot
a_2}{a_2} + \frac{m n}{a_3^2} &=& \kappa T_{3}^{3}, \label{33bvi}\\
\frac{\dot a_1}{a_1}\frac{\dot a_2}{a_2} +\frac{\dot a_2}{a_2}
\frac{\dot a_3}{a_3} + \frac{\dot a_3}{a_3}\frac{\dot a_1}{a_1} -
\frac{m^2 - m n + n^2}{a_3^2} &=&
\kappa T_{0}^{0}, \label{00bvi}\\
m \frac{\dot a_1}{a_1} - n \frac{\dot a_2}{a_2} - (m - n) \frac{\dot
a_3}{a_3} &=& \kappa T_{3}^{0}. \label{03bvi}
\end{eqnarray}
\end{subequations}
Taking into account that we are dealing with massless spinor field,
from \eqref{t00s} and \eqref{t11s} one finds $T_0^0 =  - F$,\,\, $
T_1^1 = T_2^2 = T_3^3 = S \frac{dF}{dS} - F$ and $T_{3}^{0} = 0$.
Then the Eq. \eqref{03bvi} immediately gives
\begin{equation}
\Bigl(\frac{a_1}{a_3}\Bigr)^m = {\cal N}_1
\Bigl(\frac{a_2}{a_3}\Bigr)^n, \quad {\cal N}_1 = {\rm const.}.
\label{abcrel}
\end{equation}

Let us now define
\begin{eqnarray}
V =  a_1 a_2 a_3. \label{vdef}
\end{eqnarray}

Summation of \eqref{11bvi}, \eqref{22bvi}, \eqref{33bvi} and three
times \eqref{00bvi} gives

\begin{eqnarray}
\frac{\ddot V}{V} = 2 \frac{m^2 - mn + n^2}{a_3^2} + \frac{3
\kappa}{2} \Bigl[S \frac{dF}{dS} - 2F\Bigr]. \label{vdefeq}
\end{eqnarray}

Before solving this equation let us write $S$ in terms of $v$. The
spinor field equations in this case read
\begin{subequations}
\label{spinv}
\begin{eqnarray}
\bg^0\Bigl(\dot \psi + \frac{\dot V}{2 V} \psi\Bigr)
-\frac{m-n}{2 a_3} \bg^3 \psi - i\frac{dF}{dS} \psi &=& 0, \\
\Bigl(\dot{\bp} + \frac{\dot v}{2 v}\bp\Bigr)\bg^0 -\frac{m-n}{2
a_3} \bp \bg^3  + i\frac{dF}{dS} \bp &=& 0.
\end{eqnarray}
\end{subequations}
From \eqref{spinv} one finds
\begin{equation}
\frac{d (V S)}{d t}= 0, \label{Svid}
\end{equation}
which gives
\begin{equation}
S = \frac{C_0}{V}, \quad C_0 = {\rm const.} \label{Svi}
\end{equation}

Thus we see that the equation for defining $V$ explicitly depends on
$a_3$. Here we assume that the expansion $\theta$ is proportional to
the eigenvalue $\sigma_1^1$ of shear tensor $\sigma_\mu^\nu$. In a
comoving system of reference with $U^\mu = (1,\,0,\,0,\,0)$ for
Bianchi type-VI metric we find
\begin{equation}
\theta = U^{\mu}_{;\mu} =  \frac{\dot a_1}{a_1} + \frac{\dot
a_2}{a_2} + \frac{\dot a_3}{a_3}. \label{expan}
\end{equation}
From
\begin{equation}
\sigma_{\mu \nu} = \frac{1}{2}\bigl(U_{\mu;\rho} P^{\rho}_{\nu} +
U_{\nu;\rho} P^{\rho}_{\mu}\bigr) - \frac{1}{3} \theta P_{\mu\nu},
\label{sigma}
\end{equation}
we find
\begin{equation}
\sigma_{1}^{1} = -\frac{1}{3}\bigl(-2\frac{\dot a_1}{a_1} +
\frac{\dot a_2}{a_2} + \frac{\dot a_3}{a_3}\bigr).\label{sigma11}
\end{equation}
As one sees, $S$, $\theta$ and $\sigma_1^1$ (it is true for
$\sigma_2^2$ and $\sigma_3^3$ as well) do not depend on $m$ and $n$.
Hence, they will remain unaltered for other Bianchi models following
from \eqref{bvi}. Now setting
\begin{equation}
\sigma_1^1 = q_1 \theta, \label{prop}
\end{equation}
where $q_1$ is a constant, one finds
\begin{equation}
a_1 = (a_2 a_3)^{{\cal N}_2}, \quad {\cal N}_2 = (1 + 3 q_1)/(2- 3
q_1). \label{abcrel1}
\end{equation}
Inserting this into \eqref{abcrel} and using \eqref{vdef} one
finally finds:
\begin{subequations}
\label{a123vi}
\begin{eqnarray}
a_1 &=& V^{{\cal N}_2/(1 + {\cal N}_2)}, \label{a1vi}\\
a_2 &=& {{\cal N}_1}^{-1/(2n - m)} V^{(m - n -
{\cal N}_2 m)/({\cal N}_2 + 1)(m - 2n)}, \label{a2vi}\\
a_3 &=& {{\cal N}_1}^{1/(2n - m)} V^{({\cal N}_2 m - n)/({\cal N}_2
+ 1)(m - 2n)}. \label{a3vi}
\end{eqnarray}
\end{subequations}

Let us note that perfect fluid satisfying the barotropic equation of
state, as well as Chaplygin gas are described by a massless spinor
field Lagrangian. Moreover, $S$, as well as $F(S)$ are the functions
of $V$. Therefore, we can rewrite the equations for defining $V$ as

\begin{eqnarray}
\ddot V = 2 \frac{m^2 - mn + n^2}{{\cal N}_1^{2/(2n - m)}} V^{q_2}+
{\cal D} (V), \quad {\cal D} (V) = \frac{3 \kappa}{2} \Bigl[S
\frac{dF}{dS} - 2F\Bigr]V, \label{vdefeqvi}
\end{eqnarray}
with
$$q_2 =
\frac{2(n-{\cal N}_2 m)}{(1 + {\cal N}_2)(m - 2n)} + 1 = \frac{1}{3}
- \frac{2q_1 (m + n)}{m - 2n}.$$

In case of spinor field nonlinearity given by \eqref{lspin1},  Eq.
\eqref{vdefeqvi} takes the form
\begin{eqnarray}
\ddot V = 2 \frac{m^2 - mn + n^2}{{\cal N}_1^{2/(2n - m)}} V^{q_2}+
\frac{3 \kappa \nu C_0^{1+W} (1-W)}{2} V^{-W},  \label{vdefeqvi1}
\end{eqnarray}
with the solution in quadrature
\begin{equation}
\int\frac{dV}{\sqrt{q_3 V^{q_2 +1} +3 \kappa \nu C_0^{1+W} V^{1-W} +
C_1}} = t + t_0, \quad q_3 =  \frac{4(m^2 - mn + n^2)}{(q_2 +
1){\cal N}_1^{2/(2n - m)}}. \label{Vquadq}
\end{equation}

Here $C_1$ is some integration constant.

Let us consider the case when the spinor field is given by the
Lagrangian \eqref{lspin2}. The equation for $V$ now reads
\begin{equation}
\ddot V = 2 \frac{m^2 - mn + n^2}{{\cal N}_1^{2/(2n - m)}} V^{q_2} +
\frac{3\kappa}{2} \Biggl[ \bigl(AV^{1+\gamma} + \lambda
C_0^{1+\gamma}\bigr)^{1/(1+\gamma)} + \frac{A
V^{1+\gamma}}{\bigl(AV^{1+\gamma} + \lambda
C_0^{1+\gamma}\bigr)^{\gamma/(1+\gamma)}}\Biggr], \label{vdefeqvi2}
\end{equation}
with the solution
\begin{equation}
\int \frac{dV}{\sqrt{C_1 + q_3 V^{q_2 +1} + 3 \kappa V
\bigl(AV^{1+\gamma} + \lambda C_0^{1+\gamma}\bigr)^{1/(1+\gamma)}}}
= t + t_0, \quad C_1 = {\rm const}. \quad t_0 = {\rm const}.
\label{Vquadch}
\end{equation}
Inserting $\gamma = 1$ we come to the result obtained in (Saha,
2005).

Finally we consider the case with modified quintessence. In this
case for $V$ we find
\begin{eqnarray}
\ddot V = 2 \frac{m^2 - mn + n^2}{{\cal N}_1^{2/(2n - m)}} V^{q_2}+
\frac{3\kappa}{2} \Bigl[\nu C_0^{1 + W} (1 - W) V^{-W} + 2W \ve_{\rm
cr}V/(1 + W)\Bigr], \label{vdefeqvi1mq}
\end{eqnarray}
with the solution in quadrature
\begin{equation}
\int\frac{dV}{\sqrt{q_3 V^{q_2 +1} +3 \kappa  \bigl[\nu C_0^{1 + W}
V^{1 - W} + W\ve_{\rm cr}V^2/(1 + W)\bigr] + C_1}} = t + t_0.
\label{Vquadmq}
\end{equation}
Recalling that in case of modified quintessence $W \in  (-1,\,0)$.
So the model allows cyclic mode of expansion, only when $q_2 < 1$.

\subsubsection{Bianchi type-VI$_0$ anisotropic cosmological model}

Setting $m = n$ from \eqref{bvi} we get Bianchi type-VI$_0$
cosmological model given by \eqref{bvi0}. In this case equation
\eqref{03bvi} gives
\begin{equation}
\frac{\dot a_1}{a_1} -  \frac{\dot a_2}{a_2} = 0, \label{03bvi0}
\end{equation}
with the solution
\begin{equation}
a_2 = {\cal N}_1 a_1, \label{a12bvi0}
\end{equation}
Assuming that $\sigma_1^1 \propto \theta$, on account of
\eqref{abcrel1} in this case we find

\begin{subequations}
\label{a123vi0}
\begin{eqnarray}
a_1 &=& V^{{\cal N}_2/(1 + {\cal N}_2)}, \label{a1vi0}\\
a_2 &=& {\cal N}_1  V^{{\cal N}_2/(1 + {\cal N}_2)}, \label{a2vi0}\\
a_3 &=& \frac{1}{{\cal N}_1} V^{(1 - {\cal N}_2)/(1 + {\cal N}_2)}.
\label{a3vi0}
\end{eqnarray}
\end{subequations}
The equation for $V$ in this case reads
\begin{eqnarray}
\ddot V = 2 m^2{\cal N}_1^2 V^{q_2} + {\cal D} (V).
\label{vdefeqvi0}
\end{eqnarray}
In this case for VI$_0$ we find the solution in quadrature analogous
to \eqref{Vquadq}, \eqref{Vquadch} and \eqref{Vquadmq}, for
quintessence, Chaplygin gas and modified quintessence, respectively,
with
$$q_2 = \frac{2({\cal N}_2 - 1)}{{\cal
N}_2 + 1} + 1 = \frac{1}{3} + 4 q_1, \qquad  q_3 = \frac{4 m^2 {\cal
N}_1^2}{q_2 + 1} =  \frac{3 m^2 {\cal N}_1^2}{1 + 3 q_1}.$$

As we see, in presence of a modified quintessence, BVI$_0$
cosmological model allows cyclic mode of evolution if $q_1 < 1/6$.
In case of $q_1 > 1/6$ we find the model is not bound from above as
is seen from figure \ref{BVI0mq}.

\subsubsection{Bianchi type-V anisotropic cosmological model}

Setting $m = - n$ from \eqref{bvi} we get Bianchi type-V
cosmological model given by \eqref{bv}. In this case equation
\eqref{03bvi} gives
\begin{equation}
\frac{\dot a_1}{a_1} +  \frac{\dot a_2}{a_2} - 2\frac{\dot
a_3}{a_3}= 0, \label{03bv}
\end{equation}
with the solution
\begin{equation}
a_1 a_2 = {\cal N}_1 a_3^2, \label{a12bv}
\end{equation}
Assuming that $\sigma_1^1 \propto \theta$, on account of
\eqref{abcrel1} in this case we find

\begin{subequations}
\label{a123v}
\begin{eqnarray}
a_1 &=& V^{{\cal N}_2/(1 + {\cal N}_2)}, \label{a1v}\\
a_2 &=& {\cal N}_1^{1/3}  V^{(2 - {\cal N}_2)/3(1 + {\cal N}_2)}, \label{a2v}\\
a_3 &=& \frac{1}{{\cal N}_1^{1/3}} V^{1/3}. \label{a3v}
\end{eqnarray}
\end{subequations}
The equation for $v$ in this case reads
\begin{eqnarray}
\ddot V = 6 m^2 {\cal N}_1^{2/3}  V^{q_2} + {\cal D} (V).
\label{vdefeqv}
\end{eqnarray}

In this case for V we find the solution in quadrature analogous to
\eqref{Vquadq}, \eqref{Vquadch} and \eqref{Vquadmq}, for
quintessence, Chaplygin gas and modified quintessence, respectively,
with
$$q_2 = \frac{1}{3}, \qquad  q_3 = \frac{12 m^2 {\cal N}_1^{2/3}}{q_2 + 1} = 9 m^2 {\cal N}_1^{2/3}.$$

We can conclude that in presence of modified quintessence BV model
always undergoes a cyclic mode of expansion.

\subsubsection{Bianchi type-III anisotropic cosmological model}

Setting $n = 0$ from \eqref{bvi} we get Bianchi type-VI$_0$
cosmological model given by \eqref{biii}. In this case equation
\eqref{03bvi} gives
\begin{equation}
\frac{\dot a_1}{a_1} -  \frac{\dot a_3}{a_3} = 0, \label{03biii}
\end{equation}
with the solution
\begin{equation}
a_3 = {\cal N}_1 a_1, \label{a12biii}
\end{equation}
Assuming that $\sigma_1^1 \propto \theta$, on account of
\eqref{abcrel1} in this case we find

\begin{subequations}
\label{a123biii}
\begin{eqnarray}
a_1 &=& V^{{\cal N}_2/(1 + {\cal N}_2)}, \label{a1biii}\\
a_2 &=& \frac{1}{{\cal N}_1}  V^{(1 - {\cal N}_2)/(1 + {\cal N}_2)}, \label{a2biii}\\
a_3 &=& {\cal N}_1 V^{{\cal N}_2/(1 + {\cal N}_2)}. \label{a3biii}
\end{eqnarray}
\end{subequations}
The equation for $V$ in this case reads
\begin{eqnarray}
\ddot V = 2 \frac{m^2}{{\cal N}_1^2}  V^{q_2} + {\cal D} (V).
\label{vdefeqbiii}
\end{eqnarray}

In this case for V we find the solution in quadrature analogous to
\eqref{Vquadq}, \eqref{Vquadch} and \eqref{Vquadmq}, for
quintessence, Chaplygin gas and modified quintessence, respectively,
with
$$q_2 = -\frac{2{\cal N}_2}{{\cal
N}_2 + 1} + 1 = \frac{1}{3} - 2 q_1, \qquad  q_3 = \frac{4 m^2}{(q_2
+ 1){\cal N}_1^2} = \frac{6 m^2}{(2 - 3 q_1){\cal N}_1^2}.$$

As we see, in presence of a modified quintessence, BVI$_0$
cosmological model allows cyclic mode of evolution if $q_1 > - 1/3$.

\myfigures{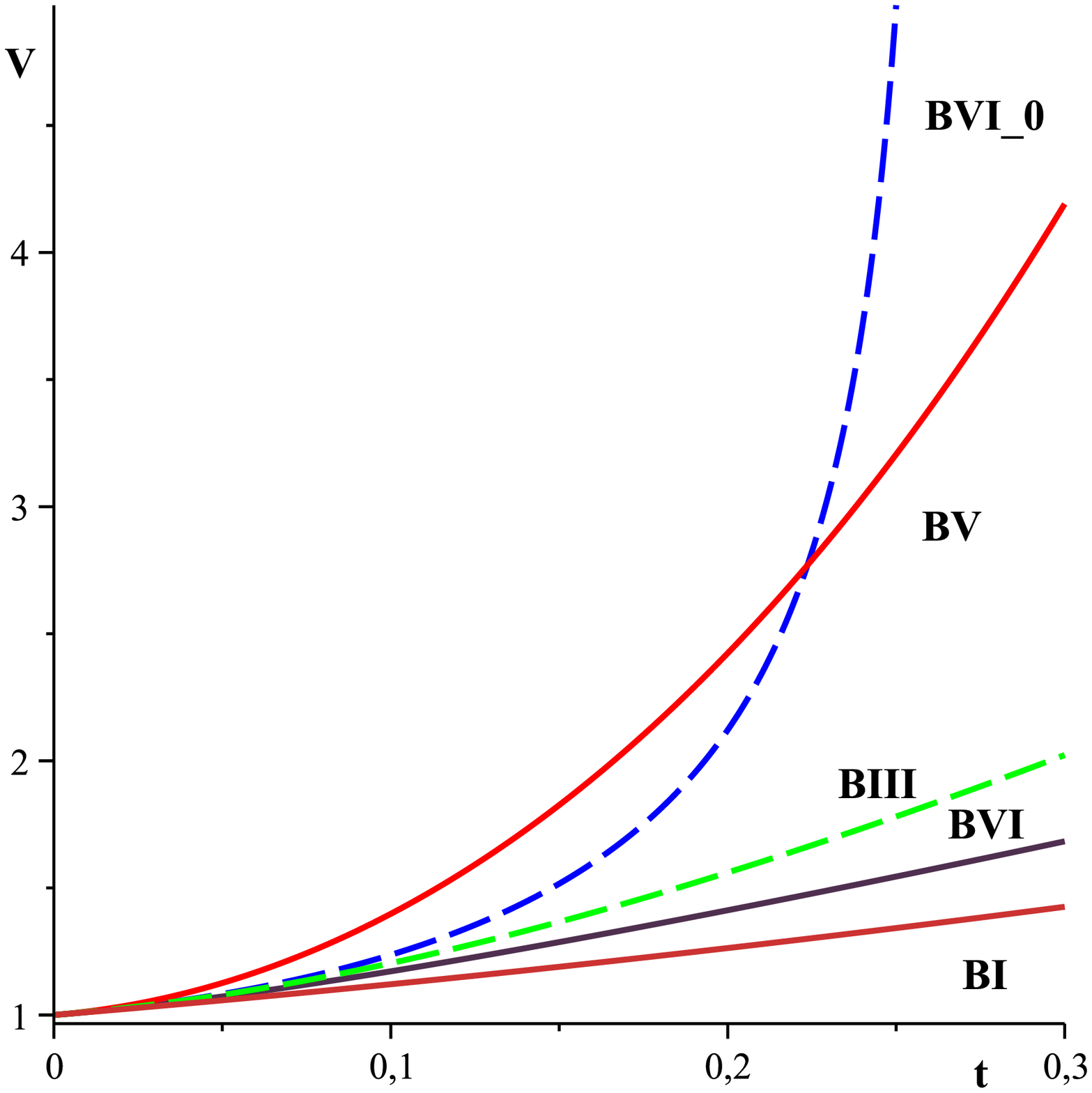}{0.45}{Evolution of the Universe filled with
quintessence for different Bianchi models}
{0.45}{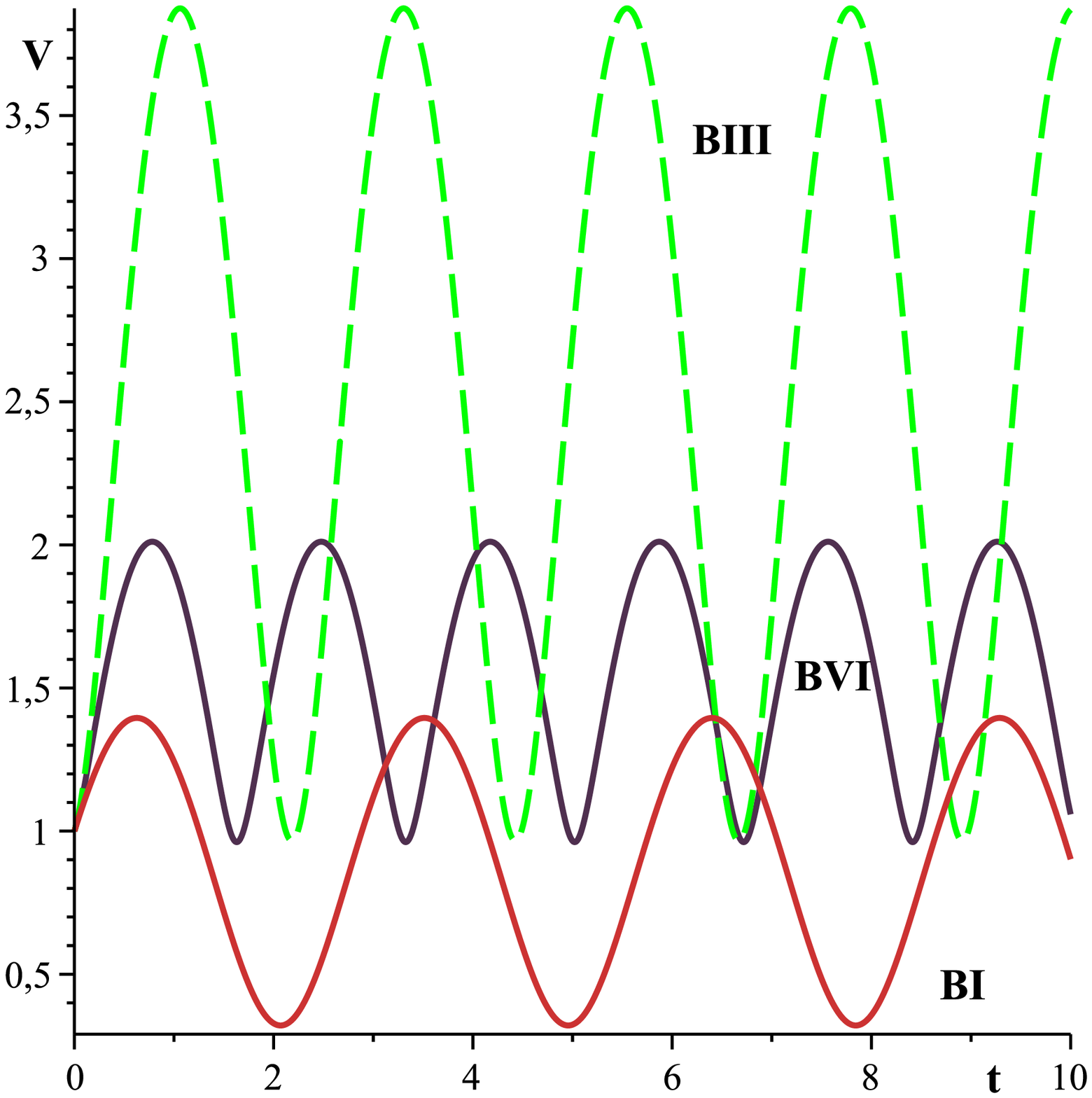}{0.45}{Evolution of the Universe filled with
modified quintessence for different Bianchi models}{0.45}

\myfigures{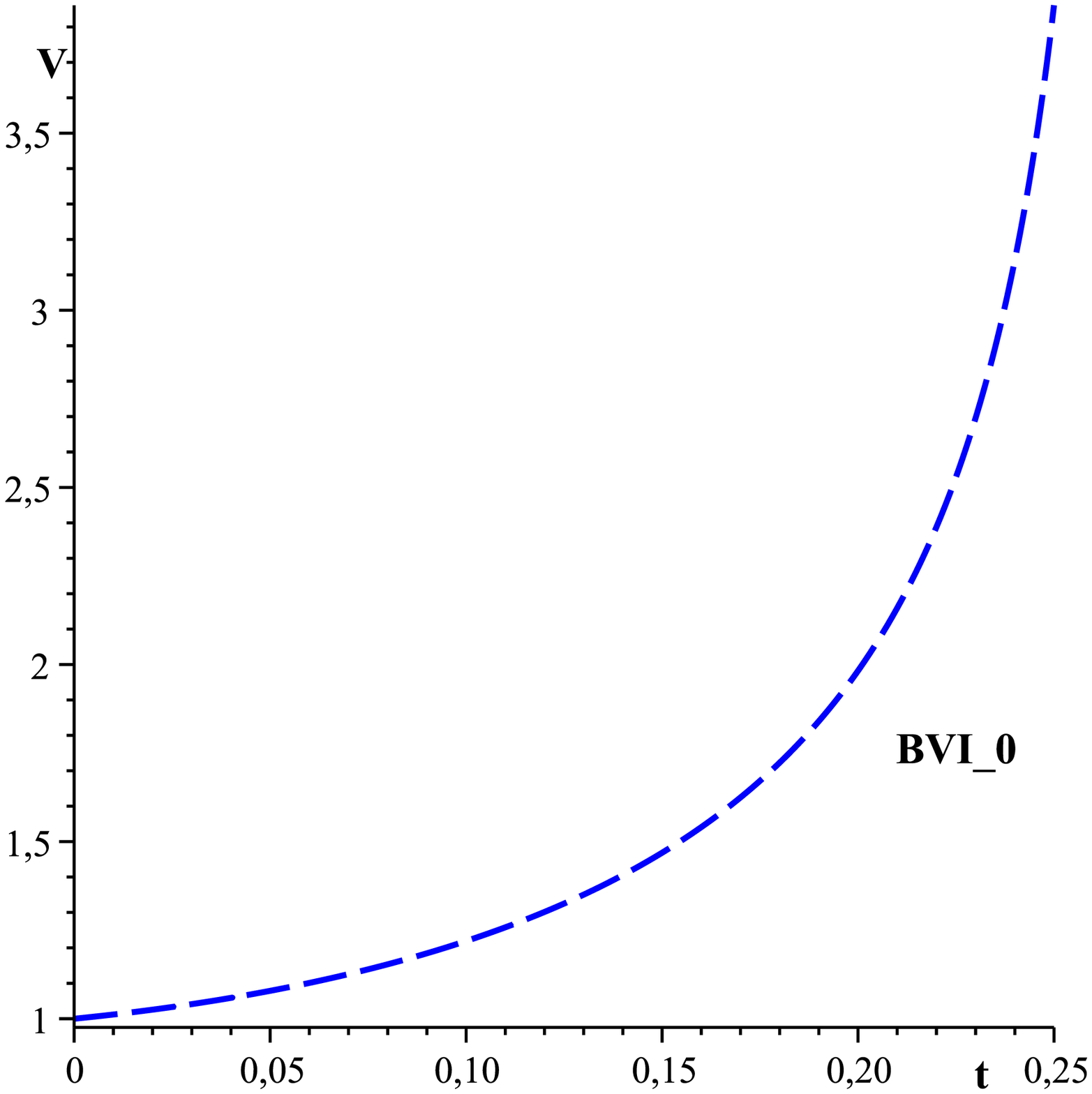}{0.45}{Evolution of the Universe filled with
modified quintessence for Bianchi type-VI$_0$ model}
{0.45}{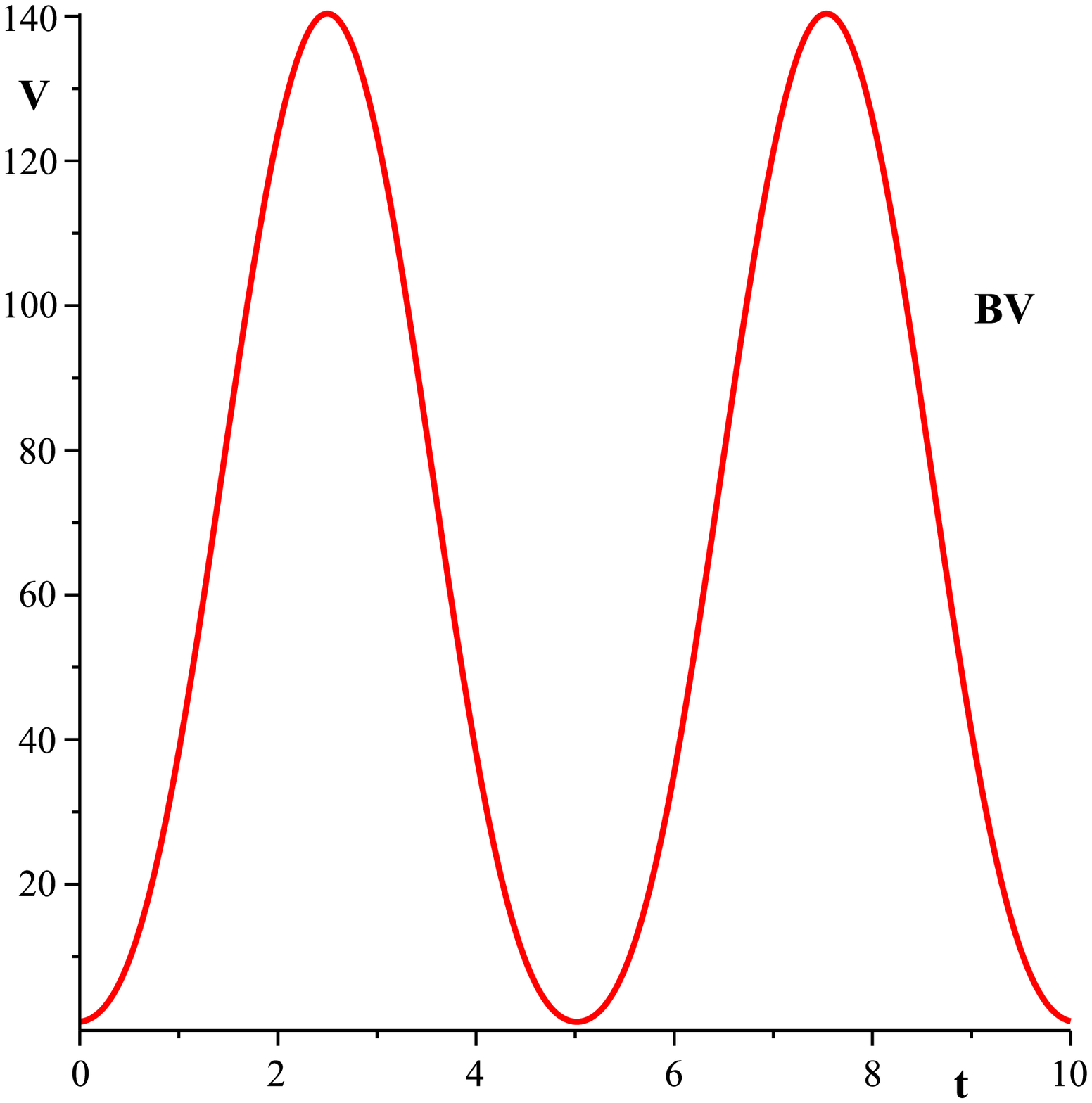}{0.45}{Evolution of the Universe filled with modified
quintessence for Bianchi type-V model}{0.45}

In the figures \ref{BVI-Iq}, \ref{BVI-III-Imq}, \ref{BVI0mq} and
\ref{BVmq} we have plotted the evolution of $V$ for different
Bianchi models filled with quintessence and modified quintessence.

\subsubsection{Bianchi type-I anisotropic cosmological model}

Bianchi type-I (BI) model is the simplest anisotropic cosmological
model and gives an excellent scope to take into account the initial
anisotropy of the Universe. Given the importance of BI model to
study the effects of initial anisotropy in the evolution of he
Universe, we study this models in details. Unlike the models
considered previously, the system of Einstein's equations for BI
model does not contain off-diagonal component \eqref{03bvi},
explicitly relating metric functions between themselves. The system
of Einstein equations in this case reads
\begin{subequations}
\label{BIE}
\begin{eqnarray}
\frac{\ddot a_2}{a_2} +\frac{\ddot a_3}{a_3} + \frac{\dot
a_2}{a_2}\frac{\dot a_3}{a_3}&=&  \kappa T_{1}^{1},\label{11bi}\\
\frac{\ddot a_3}{a_3} +\frac{\ddot a_1}{a_1} + \frac{\dot
a_3}{a_3}\frac{\dot a_1}{a_1}&=& \kappa T_{2}^{2},\label{22bi}\\
\frac{\ddot a_1}{a_1} +\frac{\ddot a_2}{a_2} + \frac{\dot
a_1}{a_1}\frac{\dot a_2}{a_2}&=&  \kappa T_{3}^{3},\label{33bi}\\
\frac{\dot a_1}{a_1}\frac{\dot a_2}{a_2} +\frac{\dot
a_2}{a_2}\frac{\dot a_3}{a_3} +\frac{\dot a_3}{a_3}\frac{\dot
a_1}{a_1}&=& \kappa T_{0}^{0}. \label{00bi}
\end{eqnarray}
\end{subequations}
Solving the Einstein equation for the metric functions one finds
(Saha, 2001a)
\begin{eqnarray}
a_i = D_i V^{1/3} \exp{\Bigl(X_i \int \frac{dt}{V}\Bigr)}, \quad
\prod_{i=1}^{3} D_i = 1, \quad     \sum_{i=1}^{3} X_i = 0,
\label{metricf}
\end{eqnarray}
with $D_i$ and $X_i$ being the integration constants.

The equation for $v$ in this case takes the form (Saha, 2001a)
\begin{eqnarray}
\ddot V = {\cal D} (V). \label{detvbi}
\end{eqnarray}
In case of \eqref{lspin1}  Eq. \eqref{detvbi} takes the form
\begin{equation}
\ddot V = (3/2) \kappa \nu C_0^{1+W} (1-W) V^{-W}
\end{equation}
with the solution in quadrature
\begin{equation}
\int\frac{dV}{\sqrt{3 \kappa \nu C_0^{1+W} V^{1-W} + C_1}} = t +
t_0.
\end{equation}
Here $C_1$ and $t_0$ are the integration constants.

\myfigures{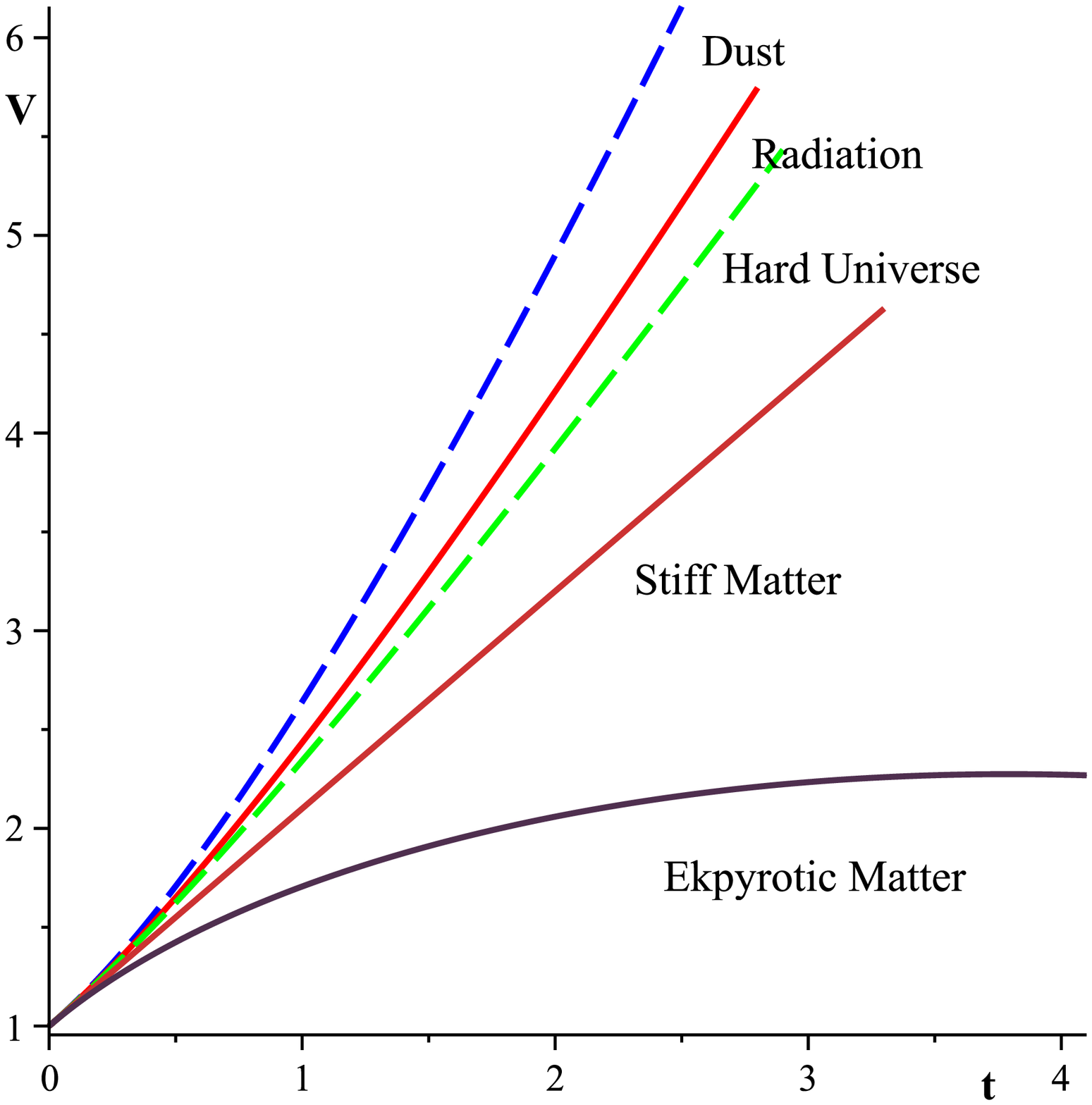}{0.45}{Evolution of the Universe filled with
perfect fluid.} {0.45}{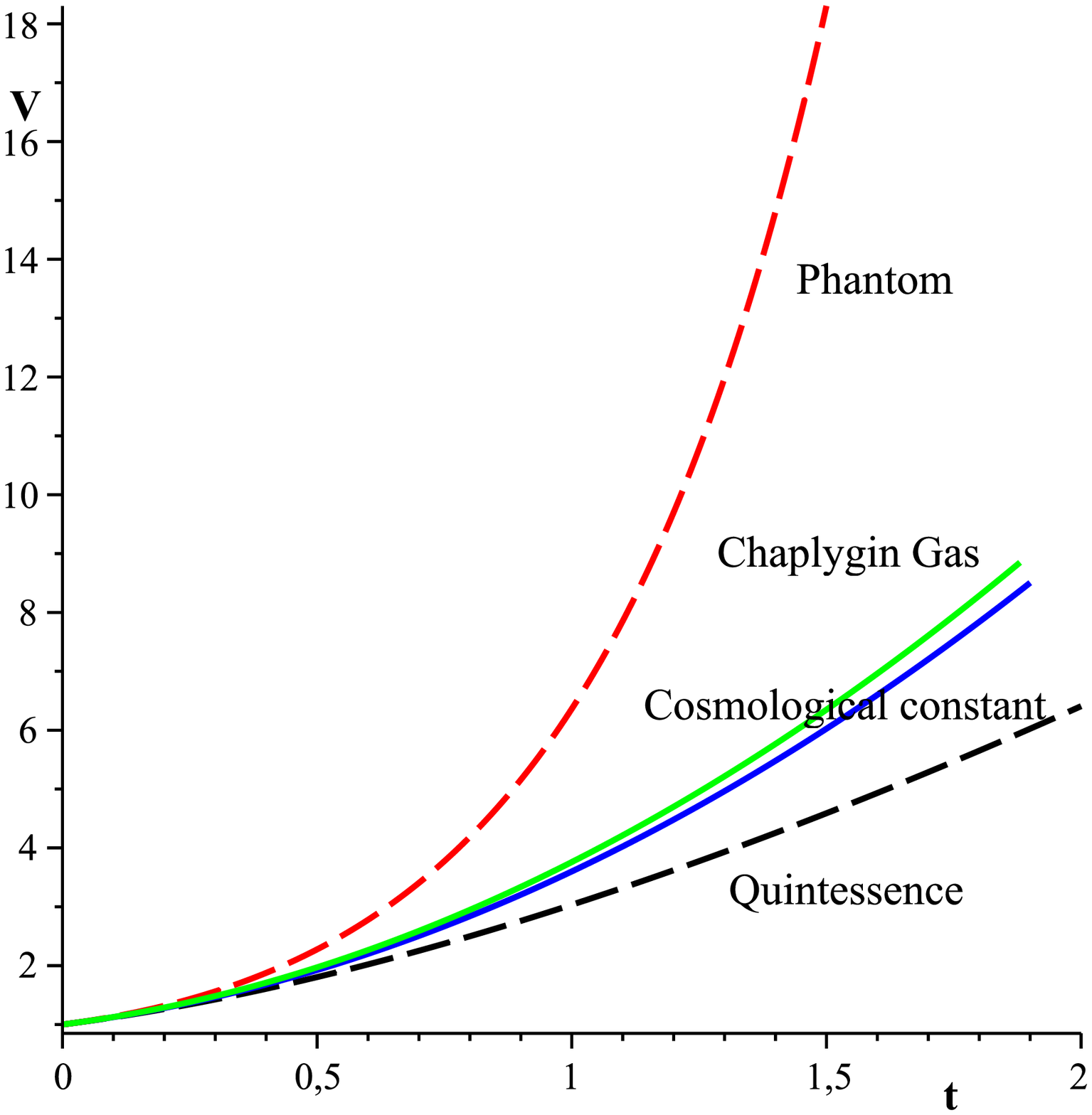}{0.45}{Evolution of the Universe
filled with dark energy.}{0.45}

In the Figs. \ref{spinpf_pf1} and \ref{spinpf_de1} we have plotted
the evolution of the Universe defined by the nonlinear spinor field
corresponding to perfect fluid and dark energy (Saha, 2010b).

Let us consider the case when the spinor field is given by the
Lagrangian \eqref{lspin2}. The equation for $V$ now reads
\begin{equation}
\ddot V = (3/2) \kappa \Biggl[ \bigl(AV^{1+\gamma} + \lambda
C_0^{1+\gamma}\bigr)^{1/(1+\gamma)} + A
V^{1+\gamma}/\bigl(AV^{1+\gamma} + \lambda
C_0^{1+\gamma}\bigr)^{\gamma/(1+\gamma)}\Biggr],
\end{equation}
with the solution
\begin{equation}
\int \frac{dV}{\sqrt{C_1 + 3 \kappa V \bigl(AV^{1+\gamma} + \lambda
C_0^{1+\gamma}\bigr)^{1/(1+\gamma)}}} = t + t_0, \quad C_1 = {\rm
const}. \quad t_0 = {\rm const}.
\end{equation}
Inserting $\gamma = 1$ we come to the result obtained in (Saha,
2005).

Finally we consider the case with modified quintessence. In this
case for $V$ we find
\begin{equation}
\ddot V = (3/2) \kappa \Bigl[\eta C_0^{1-W} (1+W) V^{W} - 2W
\ve_{\rm cr}V/(1-W)\Bigr],
\end{equation}
with the solution in quadrature
\begin{equation}
\int \frac{dV}{\sqrt{3 \kappa \bigl[\eta C_0^{1-W} V^{1+W} -
W\ve_{\rm cr}V^2/(1-W)\bigr]  + C_1}} = t + t_0. \label{qdmq}
\end{equation}
Here $C_1$ and $t_0$ are the integration constants. Comparing
\eqref{qdmq} with those with a negative $\Lambda$-term we see that
$\ve_{\rm cr}$ plays the role of a negative cosmological constant.

\myfigures{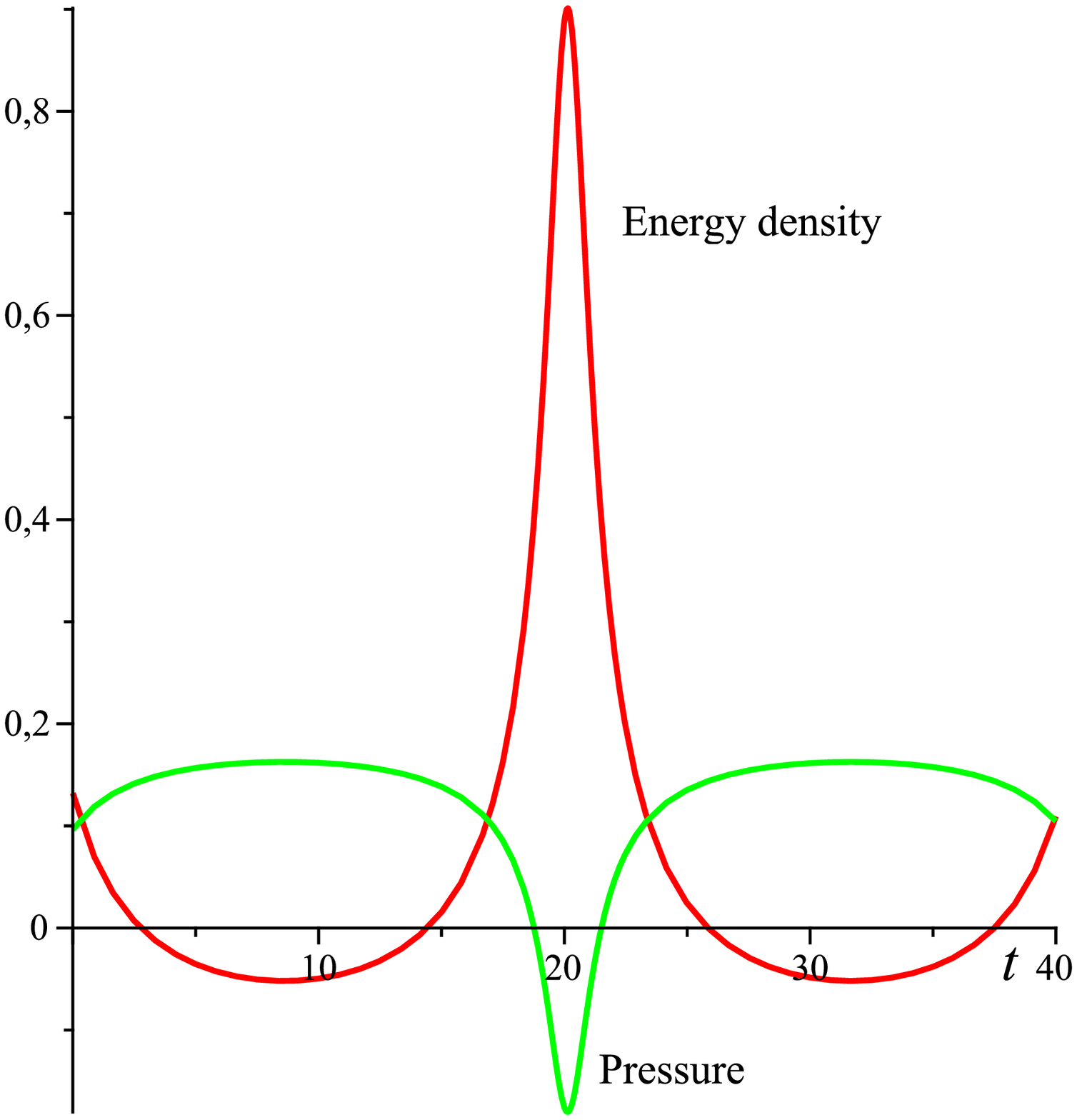}{0.45}{Dynamics of energy density and
pressure for a modified quintessence.}
{0.45}{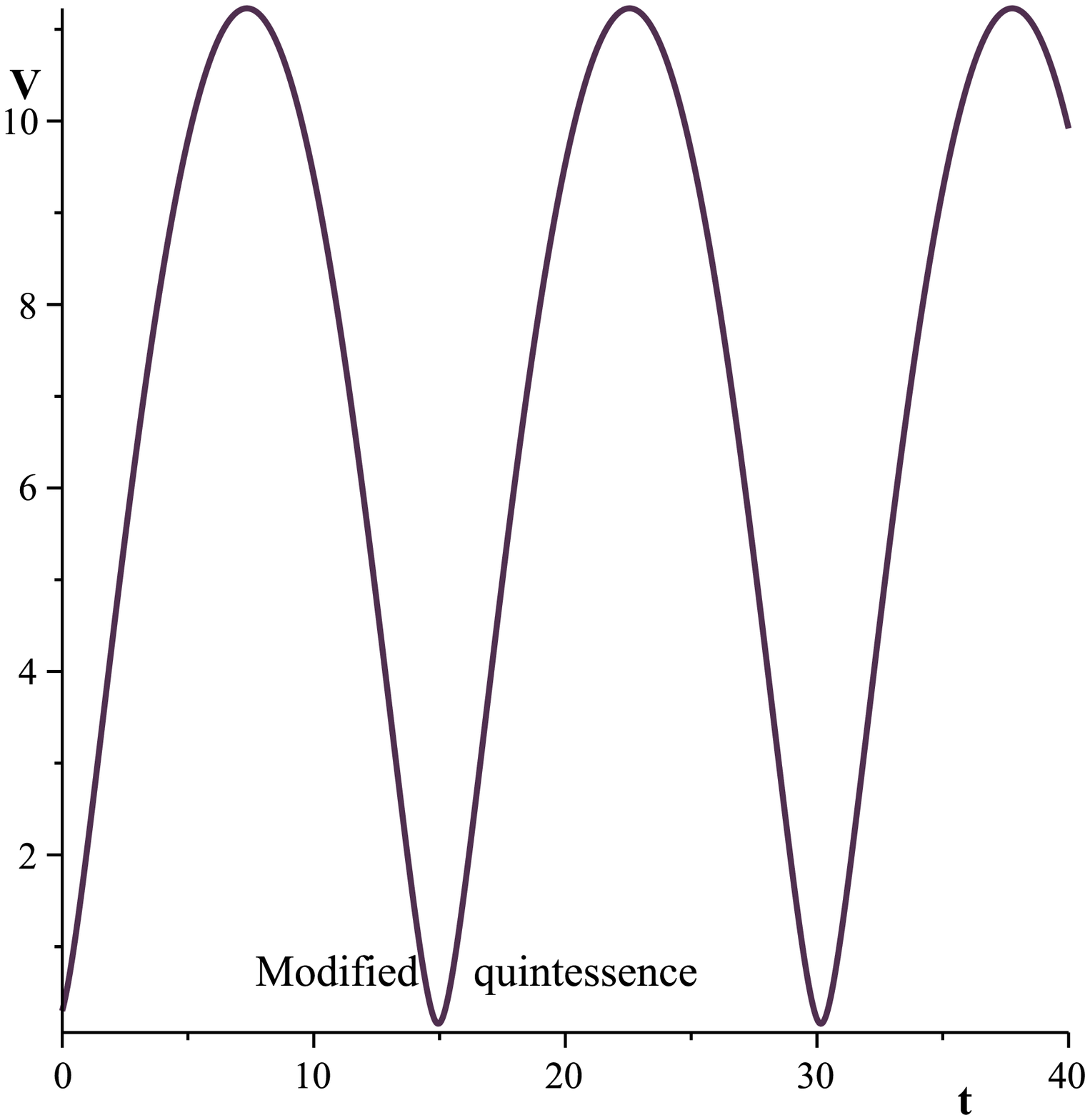}{0.45}{Evolution of the Universe filled with a
modified quintessence.}{0.45}

In the Fig. \ref{spinpf_mqep1} we have illustrated the dynamics of
energy density and pressure of a modified quintessence. In the Fig.
\ref{spinpf_mq1} the evolution of the Universe defined by the
nonlinear spinor field corresponding to a modified quintessence has
been presented. As one sees, in the case considered, acceleration
alternates with declaration. In this case the Universe can be either
singular (that ends in Big Crunch) or regular.

\subsubsection{FRW cosmological models with a spinor field}

Since our Universe is almost isotropic at large scale it would be
fitting to study evolution of the FRW model within the scope of
spinor description of matter. The Einstein equations read
corresponding to the FRW model \eqref{frw} reads
\begin{subequations}
\label{EFRW}
\begin{eqnarray}
2 \frac{\ddot a}{a} + \frac{\dot a^2}{a^2}&=& \kappa T_{1}^{1}
\label{FRW11}\\
3\frac{\dot a^2}{a^2}&=& \kappa T_{0}^{0}. \label{FRW00}
\end{eqnarray}
\end{subequations}
From the spinor field equations in this case we find
\begin{equation}
S = \frac{C_0}{a^3}, \quad C_0 = {\rm const.} \label{SFRW}
\end{equation}

In order to find the solution that satisfies both \eqref{FRW11} and
\eqref{FRW00} we rewrite \eqref{FRW11} in view of \eqref{FRW00} in
the following form:
\begin{equation}
\ddot a = \frac{\kappa}{6}\Bigl(3 T_1^1 - T_0^0\Bigr) a. \label{dda}
\end{equation}

Further we solve this equation for concrete choice of source field.

Let us consider the case of perfect fluid given by the barotropic
equation of state. In account of \eqref{t00sf}, \eqref{t11sf} and
\eqref{SFRW},  \eqref{dda} takes the form
\begin{equation}
\ddot a = \frac{\kappa \nu (1+3W)C_0^{1+W}}{2} a^{-(2+3W)},
\label{ddasf}
\end{equation}
that admits the first integral
\begin{equation}
\dot a^2 =  \frac{\kappa}{3} \nu C_0^{1+W} a^{-(1+3W)} + E_1, \quad
E_1 = {\rm const}. \label{dda1}
\end{equation}

In Fig. \ref{spinpf_pf_FRW} and \ref{spinpf_phan_FRW} we plot the
evolution of the FRW Universe for different values of $W$.

\myfigures{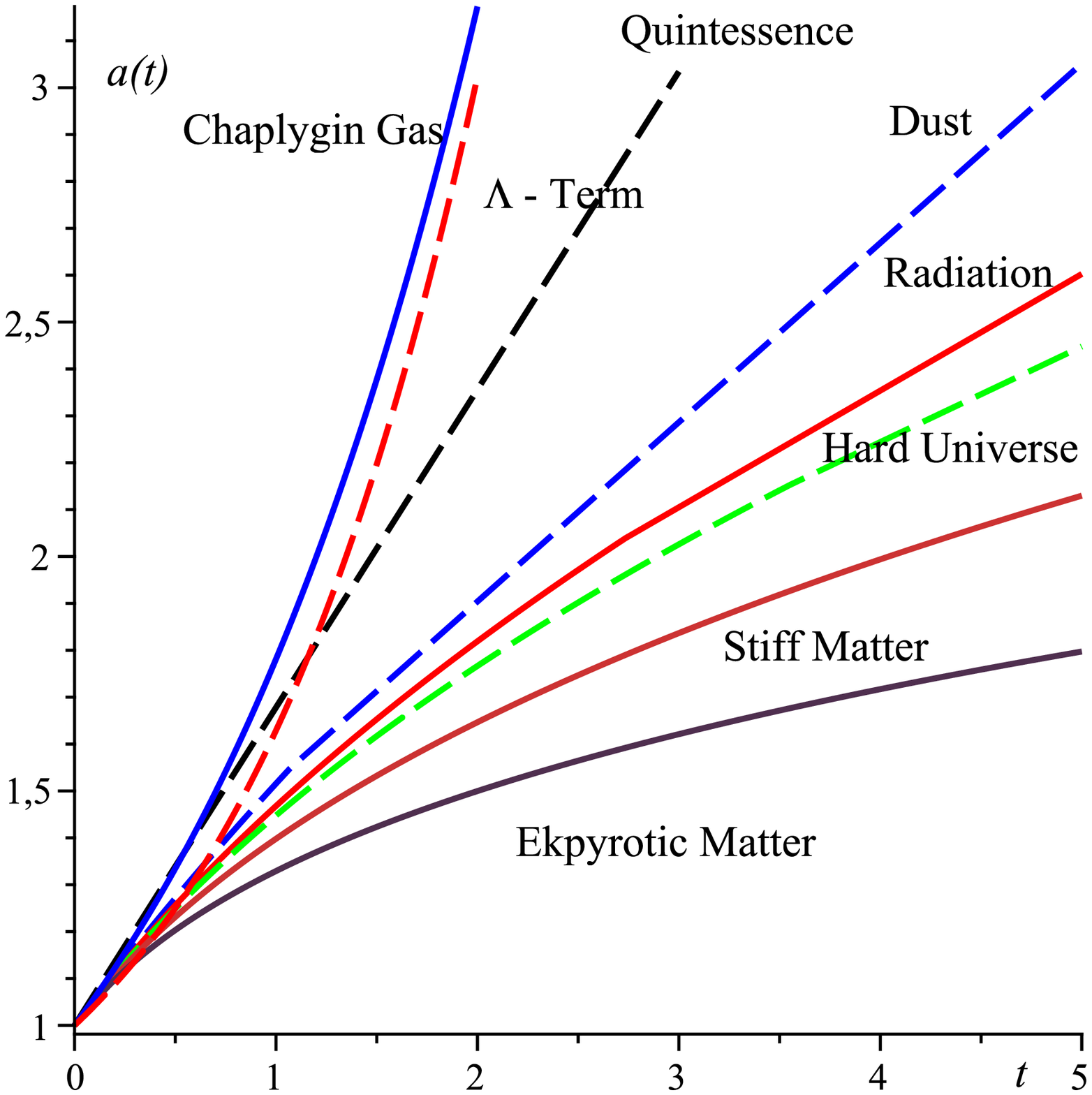}{0.45}{Evolution of the Universe filled
with perfect fluid and dark energy.}
{0.45}{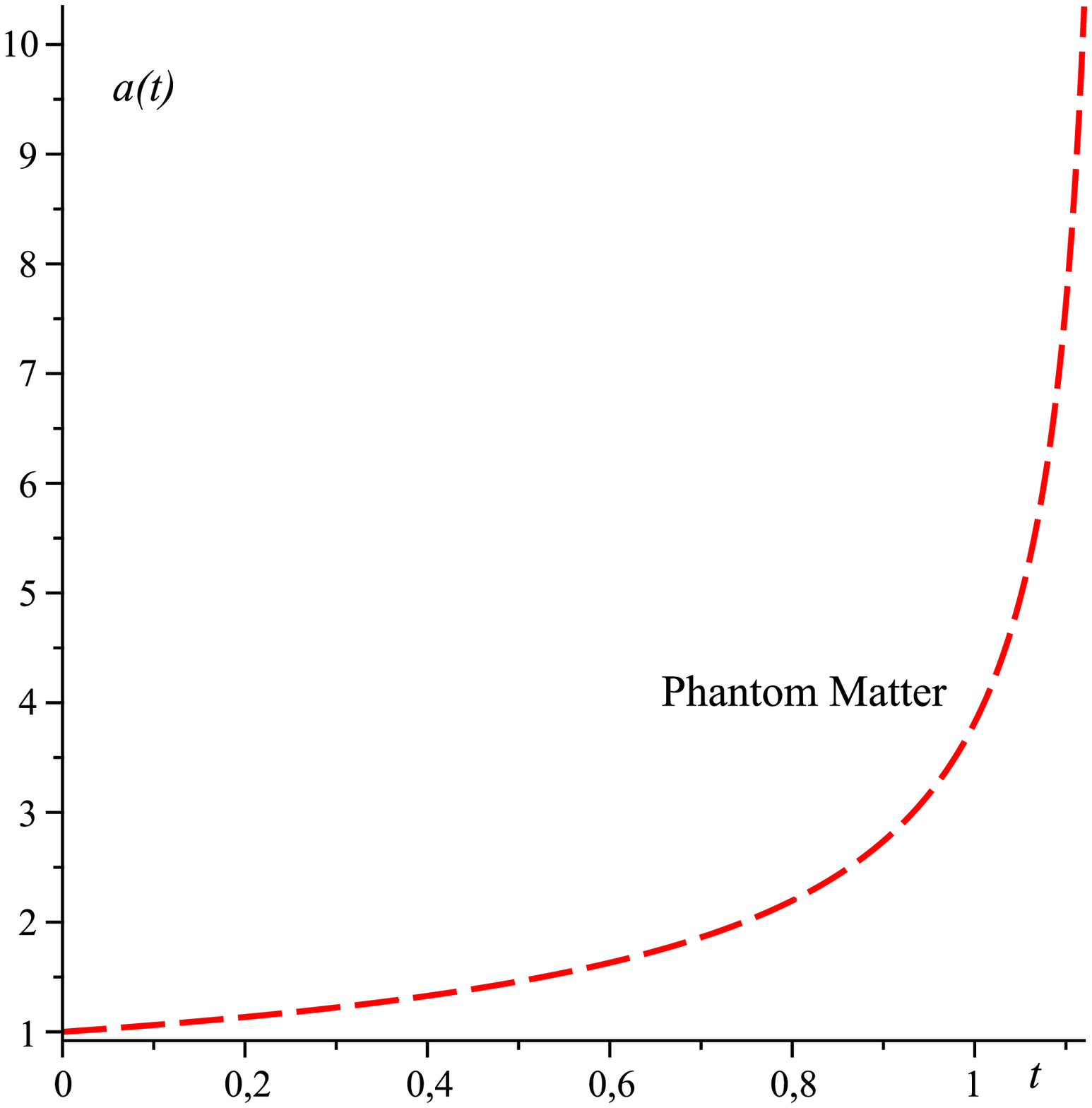}{0.45}{Evolution of the Universe filled with
phantom matter.}{0.45}

As one sees, equation \eqref{dda1} imposes no restriction on the
value of $W$. But it is not the case, when one solves \eqref{FRW00}.
Indeed, inserting $T_0^0$ from \eqref{t00sf} into \eqref{FRW00} one
finds
\begin{equation}
a = (A_1 t + C_1)^{2/3(1+W)}, \label{afrw}
\end{equation}
where $A_1 = (1+W)\sqrt{3 \kappa \nu C_0^{1+W}/4}$ and $C_1 = 3 (1 +
W) C/2$ with $C$ being some arbitrary constant. This solution
identically satisfies the equation \eqref{FRW11}. As one sees, case
with $W = -1$, cannot be realized here. In that case one has to
solve the equation \eqref{FRW00} straight forward. As far as phantom
matter ($W < -1$) is concerned, there occurs some restriction on the
value of $C$, as in this case $A_1$ is negative and for the $C_1$ to
be positive, $C$ should be negative. As one can easily verify, in
case of cosmological constant with $W = -1$ Eqn. \eqref{FRW00} gives
\begin{equation}
a  = a_0 e^{\pm \sqrt{\kappa \nu/3}\,t}. \label{FRWlambda}
\end{equation}

Inserting \eqref{edchapsp} and \eqref{prchapsp} into \eqref{dda} in
case of Chaplygin gas we have the following equation
\begin{equation}
\ddot a  = \frac{\kappa}{6}\frac{2A a^{3(1+\gamma)} - \lambda
C_0^{1+\gamma}}{a^2\Bigl(Aa^{3(1+\gamma)} + \lambda
C_0^{1+\gamma}\Bigr)^{\gamma/(1+\gamma)}}. \label{FRWchap}
\end{equation}
We solve this equation numerically. The corresponding solution has
been illustrated in Fig. \ref{spinpf_pf_FRW}.

Finally we consider the case with modified quintessence. Inserting
\eqref{edmq} and \eqref{prmq} into \eqref{dda} in this case we find
\begin{equation}
\ddot a  = \frac{\kappa}{6}\Bigr[(3W-1)\eta C_0^{1-W} a^{3W-2} -
\frac{2W}{1-W}\ve_{\rm cr} a\Bigr], \label{FRWmq}
\end{equation}
with he solution
\begin{equation}
\dot a^2 =  \frac{\kappa}{3}\Bigl[\nu C_0^{1-W} a^{3W - 1} -
\frac{W}{1-W} \ve_{\rm cr} a^2 + E_2\Bigr], \quad E_2 = {\rm const}.
\label{dda2}
\end{equation}
It can be shown that in case of modified quintessence the pressure
is sign alternating. As a result we have a cyclic mode of evolution.

\section{Singularity problem}

On the the main problems of modern cosmology is the singularity
problem. Let us study this problem within the scope of models
discussed above. As we see, the components of the spinor field and
metric functions are expressed in terms of $v$. It should be noted
that $v$ plays one of the central roles in studying the singular
space-time points of BI cosmological models. Here we describe it in
brief. In doing so let us first write the Kretschmann scalar for the
Bianchi type-VI metric. Taking into account that the metric
\eqref{bvi} possesses the following non-trivial components of
Riemann tensor:
\begin{eqnarray}
R_{\,\,\,01}^{01} &=& -\frac{\ddot a_1}{a_1}, \quad
R_{\,\,\,02}^{02} = -\frac{\ddot a_2}{a_2}, \quad
R_{\,\,\,03}^{03} = -\frac{\ddot a_3}{a_3}, \nonumber \\
R_{\,\,\,12}^{12} &=& -\frac{mn}{a_3^2} - \frac{\dot
a_1}{a_1}\frac{\dot a_2}{a_2},\quad R_{\,\,\,13}^{13} =
\frac{m^2}{a_3^2} - \frac{\dot a_3}{a_3}\frac{\dot a_1}{a_1}, \quad
R_{\,\,\,23}^{23} = \frac{n^2}{a_3^2} - \frac{\dot
a_2}{a_2}\frac{\dot a_3}{a_3}, \nonumber\\
R_{\,\,\,10}^{31} &=& \frac{m}{a_3^2} \Bigl(\frac{\dot a_1}{a_1} -
\frac{\dot a_3}{a_3} \Bigr),\quad R_{\,\,\,13}^{01} = m
\Bigl(\frac{\dot a_3}{a_3} - \frac{\dot a_1}{a_1} \Bigr),\nonumber\\
R_{\,\,\,20}^{32} &=& \frac{n}{a_3^2} \Bigl(\frac{\dot a_3}{a_3} -
\frac{\dot a_2}{a_2} \Bigr),\quad R_{\,\,\,23}^{02} = n
\Bigl(\frac{\dot a_2}{a_2} - \frac{\dot a_3}{a_3} \Bigr), \nonumber
\end{eqnarray}
for the Kretschmann scalar we find
\begin{eqnarray}
{\cal K} &=& R_{\mu\nu\alpha\beta} R^{\mu\nu\alpha\beta} =
R^{\mu\nu}_{\,\,\,\,\,\,\,\alpha\beta} R_{\,\,\,\,\,\,\,\mu\nu}^{\alpha\beta} \nonumber \\
 & = & 4\Bigl[\Bigl(\frac{\ddot a_1}{a_1}\Bigr)^2 + \Bigl(\frac{\ddot
a_2}{a_2}\Bigr)^2 + \Bigl(\frac{\ddot a_3}{a_3}\Bigr)^2 +
\Bigl(\frac{\dot a_1}{a_1}\frac{\dot a_2}{a_2}\Bigr)^2 +
\Bigl(\frac{\dot a_2}{a_2}\frac{\dot a_3}{a_3}\Bigr)^2 +
\Bigl(\frac{\dot a_3}{a_3}\frac{\dot a_1}{a_1}\Bigr)^2\Bigr]
\label{Kretsch}\\
& + & \frac{4}{a_3^4}\Bigl[(m^4 + m^2n^2 + n^4) - (m^2 + n^2){\dot
a_3}^2 - a_3^2 \Bigl(m \frac{\dot a_1}{a_1} - n \frac{\dot
a_2}{a_2}\Bigr)^2\Bigr]. \nonumber
\end{eqnarray}

The metric functions for BVI, BVI$_0$, BV and BIII can be expressed
as
\begin{equation}
a_i = {\cal B}_i V^{s_i}, \label{Vgen}
\end{equation}
which gives
\begin{subequations}
\label{singvis}
\begin{eqnarray}
\frac{\dot a_i}{a_i} &=& {\cal B}_i \frac{\dot{V}}{V},\\
\frac{\ddot a_i}{a_i} &=& {\cal B}_i \frac{\ddot{V}}{V} + {\cal B}_i
({\cal B}_i -1) \Bigl(\frac{\dot{V}}{V}\Bigr)^2.
\end{eqnarray}
\end{subequations}
The metric functions and their derivatives for the BI take the
following form:
\begin{subequations}
\label{singb1}
\begin{eqnarray}
\frac{\dot a_i}{a_i} &=& \frac{\dot{V}}{3 V} + \frac{X_i}{3V},\\
\frac{\ddot a_i}{a_i} &=& \frac{\ddot{V}}{3V} - \frac{2}{9}
\Bigl(\frac{\dot{V}}{V}\Bigr)^2 - \frac{X_i}{9}\frac{ \dot{V}}{V^2}
+ \frac{X_i^2}{9 V^2}.
\end{eqnarray}
\end{subequations}

We study the singularity on the basis of Kretschmann scalar and in
doing so we follow the criteria given in (Bronnikov {\it et al},
2004):

(i) For any finite $t$  some $a_i \to 0$. \eqref{Kretsch} shows that
if more than one scale factor becomes trivial at finite $t$, then it
is a singularity. As is seen from  \eqref{singb1} as $V \to 0$ (i)
all $a_i \to 0$ if $X_1 = X_2 = X_3 = 0$, i.e., it is a singularity;
(ii) more than one $a_i \to 0$ if more than one $X_i < 0$ and in
this case we have singularity; (iii) only one $a_i \to 0$ if only
one $X_i < 0$, i.e., the space-time can be non-singular.

Note that this criteria of singularity always fulfills at the point
where $V = 0$.

As far as  $t \to \infty$ is concerned, the corresponding asymptote
can be singular if at least one $a_i$ vanishes faster than
exponentially. As $X_1 + X_2 + X_3 = 0$, it means one or more $X_i$
is negative, and from \eqref{singvis} it follows that at least one
function $a_i$ vanishes faster than exponentially. Hence it is a
singularity.

As far as BVI, BVI$_0$, BV, BIII and FRW models are concerned, they
are all singular at a space-time point, where $V = 0$. Moreover, in
cases of BVI, BVI$_0$, BV and BIII at least two scale factor $a_i$'s
are directly related, hence there is always a possibility of more
than one one scale factor becoming trivial, thus giving rise to a
singularity at finite $t$.

Moreover, components of the spinor field as well as the physically
observable quantities such as charge, current, spin etc. constructed
from them are inverse functions of $V$ (Saha, 2001a). Hence we
conclude that within the scope of the model considered here at any
point where $V = 0$ there occurs a space-time singularity.

\section{Problem of isotropization}

Since the present-day Universe is surprisingly isotropic, it is
important to see whether our anisotropic BI model evolves into an
isotropic FRW model. Isotropization means that at large physical
times $t$, when the volume factor $V$ tends to infinity, the three
scale factors $a_i(t)$ grow at the same rate. Two wide-spread
definition of isotropization read
\begin{subequations}
\label{aniso}
\begin{eqnarray}
{\cal A} &=&  \frac{1}{3} \sum\limits_{i=1}^{3} \frac{H_i^2}{H^2} - 1  \to 0,\\
\Sigma^2 &=& \frac{1}{2}  {\cal A} H^2 \to 0.
\end{eqnarray}
\end{subequations}
Here ${\cal A}$ and $\Sigma^2$ are the average anisototropy and
shear, respectively. $H_i = \dot{a_i}/a_i$ is the directional Hubble
parameter and $H = \dot{a}/a$ average Hubble parameter, where  $a(t)
= V^{1/3}$ is the average scale factor. Here we exploit the
isotropization condition proposed proposed in Bronnikov {\it et al}
(2004):
\begin{equation}
\frac{a_i}{a}\Bigl|_{t \to \infty} \to {\rm const.} \label{isocon}
\end{equation}
Then by rescaling some of the coordinates, we can make $a_i/a \to
1$, and the metric will become manifestly isotropic at large $t$. It
can be shown that in case of BVI, BVI$_0$, BV and BIII models the
criteria \eqref{isocon} does not hold, hence the isotropization
process for these models does not take place. So we consider the BI
model and study the problem od isotropization for this model in
details.

From \eqref{metricf} we find
\begin{eqnarray}
\frac{a_i}{a} = \frac{a_i}{V^{1/3}} =  D_i \exp{\Bigl(X_i \int
\frac{dt}{\tau}\Bigr)}.\label{isocon1}
\end{eqnarray}
As is seen from \eqref{metricf} in our case $a_i / a \to D_i =$
const as $v \to \infty$. Recall that the isotropic FRW model has
same scale factor in all three directions, i.e., $a_1(t) = a_2(t) =
a_3(t) = a(t)$. So for the BI universe to evolve into a FRW one the
constants $D_i$'s are likely to be identical, i.e., $D_1 = D_2 = D_3
= 1$. Moreover, the isotropic nature of the present Universe leads
to the fact that the three other constants $X_i$ should be close to
zero as well, i.e., $|X_i| << 1$, ($i = 1,2,3$), so that $X_i \int
[v (t)]^{-1}dt \to 0$ for $t < \infty$ (for $V (t) = t^n$ with $n
> 1$ the integral tends to zero as $t \to \infty$ for any $X_i$). It
can be concluded that the spinor field Lagrangian with $W < 1$ leads
to the isotropization of the Universe as $t \to \infty$, moreover,
in case of $W < 0$ the system undergoes an earlier isotropization.

\section{Discussion}

Let us now examine what kind of advantage one gets exploiting the
spinor description of matter. We do it  within the scope of a BI
cosmological model. In doing so, we consider the case when the
Universe is filled with, say, Van-der-Waals fluid, radiation and
quintessence. In this case we have

\begin{subequations}
\label{3comp}
\begin{eqnarray}
T_0^0 &=&  \ve_{\rm v} + \ve_{\rm r} + \ve_{\rm q},\\
T_1^1 &=&  - p_{\rm v} - p_{\rm r} - p_{\rm q}.
\end{eqnarray}
\end{subequations}
To solve \eqref{vdefeqvi} one has to know $T_0^0$ and $T_1^1$ in
terms of $v$. It can be done exploiting Bianchi identity
\begin{equation}
\dot T_0^0 + \frac{\dot V}{V} \Bigl(T_0^0 - T_1^1\Bigr) = 0.
\label{bianid}
\end{equation}
The dark energy is supposed to interact with itself only, so it is
minimally coupled to the gravitational field. As a result, the
evolution equation for the energy density decouples from that of the
perfect fluid. Taking this into account and inserting \eqref{3comp}
into \eqref{bianid}, we obtain two balance equations
\begin{subequations}
\label{2eq}
\begin{eqnarray}
\dot \ve_{\rm v} + \frac{\dot V}{V} \Bigl(\ve_{\rm v} + p_{\rm
v}\Bigr) + \dot \ve_{\rm r} + \frac{\dot V}{V} \Bigl(\ve_{\rm
r} + p_{\rm r}\Bigr) &=& 0, \label{2eq1}\\
\dot \ve_{\rm q} + \frac{\dot V}{V} \Bigl(\ve_{\rm q} + p_{\rm
q}\Bigr) &=& 0. \label{2eq2}
\end{eqnarray}
\end{subequations}
In usual approach we are in trouble, as neither $\ve_{\rm v}$ nor
$\ve_{\rm r}$ cannot be expressed in terms of $v$ from \eqref{2eq1}.
But if one uses spinor description for radiation, thanks to spinor
field equation one finds
\begin{equation}
\dot \ve_{\rm r} + \frac{\dot V}{V} \Bigl(\ve_{\rm r} + p_{\rm
r}\Bigr) \equiv 0. \label{rad}
\end{equation}
As a result, in place of \eqref{2eq1} we now have
\begin{equation}
\dot \ve_{\rm v} + \frac{\dot V}{V} \Bigl(\ve_{\rm v} + p_{\rm
v}\Bigr) = 0, \label{2eq1n}
\end{equation}
which is quite computable for the given equation of state. For
Van-der-Waals fluid we use the following equation of state
\begin{equation}
p_{\rm v} = \frac{8 W_1 \ve_{\rm v}}{3 - \ve_{\rm v}} - 3 \ve_{\rm
v}^2, \label{vwf}
\end{equation}
where $W_1$ is a constant. The dark energy density $\ve_{\rm q}$ can
be expressed in terms of $V$ either solving \eqref{2eq2} or using
the spinor description. Thus using the spinor description
\eqref{t00sf} and \eqref{t11sf} for radiation and dark energy, and
defining the Hubble parameter, we find the following system of
equations
\begin{subequations}
\label{three}
\begin{eqnarray}
\dot V &=& 3 H V, \label{tau1}\\
\dot H &=& -3H^2 + \frac{\kappa}{2} \Bigl[\ve_{\rm v} -
\frac{8W_1\ve_{\rm v}}{3-\ve_{\rm v}} + 3 \ve_{\rm v}^2 + \frac{2
\ve_{\rm r0}}{3}V^{-4/3} + (1-W) \ve_{\rm q0}V^{-1 -W}\Bigr],
\label{H}\\
\dot \ve_{\rm v} &=& - 3 \Bigl[\ve_{\rm v} + \frac{8W_1 \ve_{\rm
v}}{3 - \ve_{\rm v}} - 3 \ve_{\rm v}^2\Bigr] H. \label{vwf1}
\end{eqnarray}
\end{subequations}
For simplicity, we set $\ve_{\rm r0} = 1$ and $\ve_{\rm q0} = 1$.
The equation \eqref{vwf1} can be exactly solved for $W_1 = 1/2$. In
this case we find
\begin{equation}
\frac{\ve_{\rm v}^6 (3 \ve_{\rm v} - 7)}{(\ve_{\rm v} - 1)^7} =
\frac{C_0}{V^{14}}, \quad C_0 = const.\label{vwr}
\end{equation}
As one sees, it is quite a complicate expression. In what follows,
we solve the system \eqref{three} numerically. For this purpose we
also set $W = -1/2$ for quintessence.

\myfigures{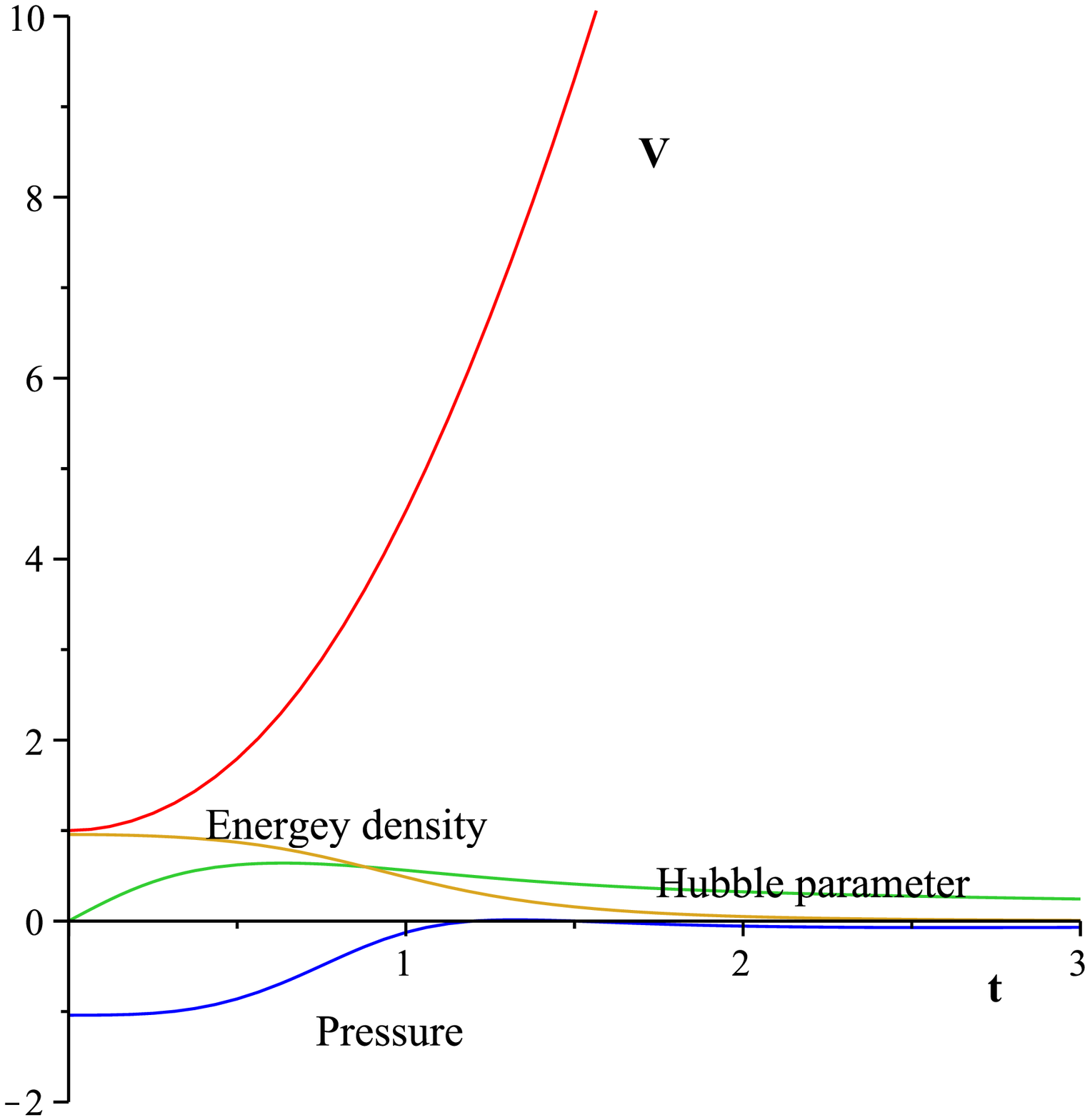}{0.45}{Evolution of the $V$, $H$ and $\ve_{\rm v}$
and $p$ for $V (0) = 1$, $H (0) = 0.0000945$ and $\ve_{\rm v} =
0.957$}{0.45}{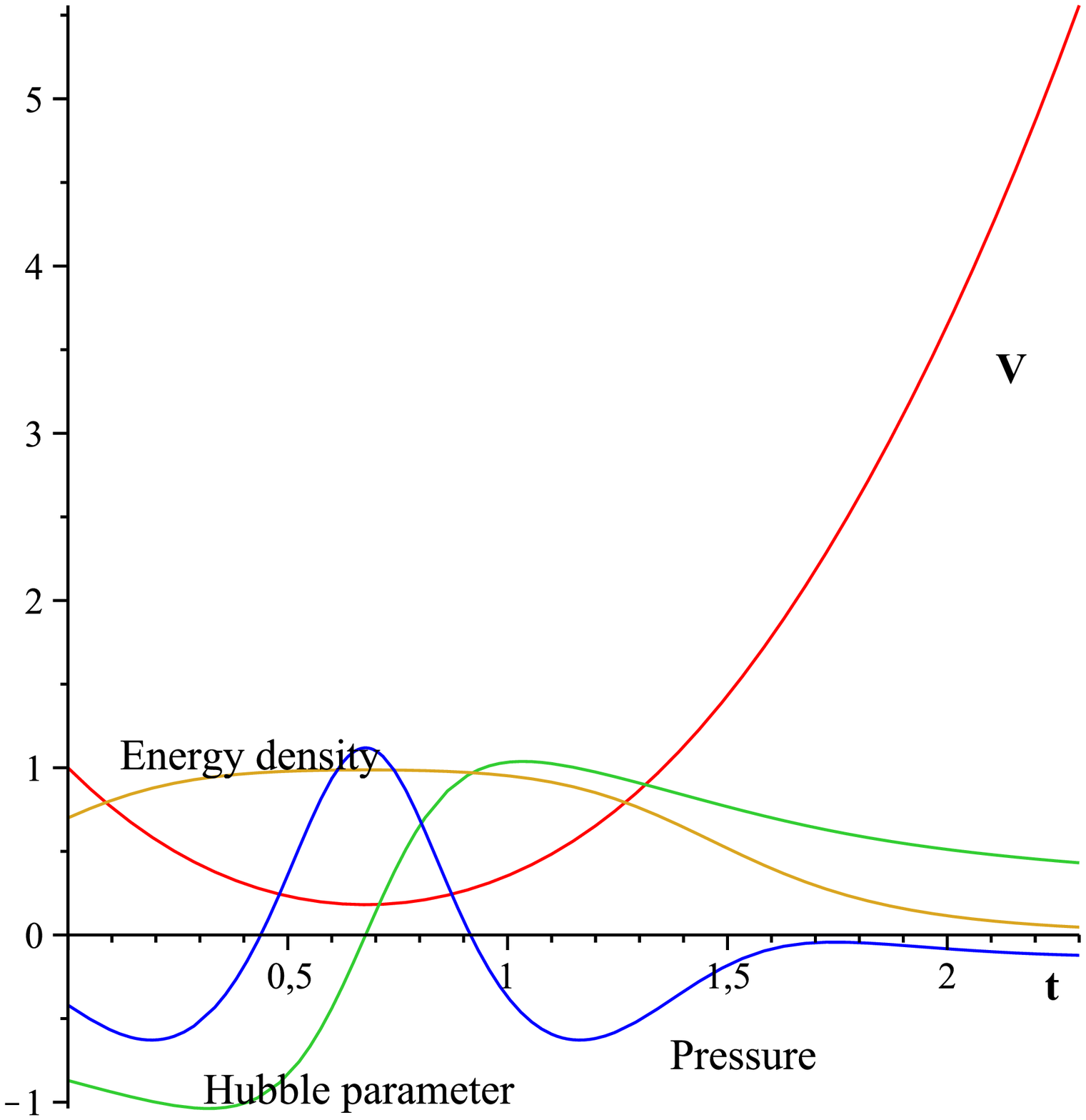}{0.45}{Evolution of the $V$, $H$, $\ve_{\rm v}$
and $p$ for $V (0) = 1$, $H (0) = -0.9$ and $\ve_{\rm v} =
0.7$}{0.45}

In the figures \ref{VEPH1} and  \ref{VEPH2} we plot the evolution of
$V$, $H$ and $\ve_{\rm v}$ and $p$ for different initial values. As
one sees, the pressure has negative value at the initial stage of
evolution, then it begin to increase, thus giving rise to
deceleration and finally becomes negative that results in the late
time acceleration of the Universe.

\section{Conclusion}

Within the framework of cosmological gravitational field equivalence
between the perfect fluid (and dark energy) and nonlinear spinor
field has been established. It is shown that different types of dark
energy can be simulated by means of a nonlinear spinor field. Using
the new description of perfect fluid or dark energy evolution of the
Universe has been studied within the scope of a BVI, BVI$_0$, BV,
BIII and BI anisotropic models as well as isotropic FRW model. The
corresponding Einstein equations have been solved. It is shown that
all the models give rise to a space-time singularity where $V$ is
trivial. Among the Bianchi models considered only the BI allows
isotropization of the initially anisotropic space-time. It is shown
that the spinor description of fluid allows one to solve the
two-fluid system without without imposing any additional condition.



\end{document}